\documentclass[aps,pre,superscriptaddress,reprint]{revtex4-1}

\usepackage[version=3]{mhchem} 
\usepackage{graphicx}
\usepackage{amssymb}
\usepackage{epstopdf}
\usepackage{epsfig}
\usepackage{color}
\usepackage{amsmath}
\usepackage{amsfonts}
\usepackage{graphicx}
\usepackage{fancybox}
\usepackage{subfigure}
\usepackage{notes2bib}

\newcommand{\rb}{\mathbf{r}}

\newcommand{\avg}[1]{\left<#1\right>}
\newcommand{\len}[1]{\left|#1\right|}
\newcommand{\brac}[1]{\left[#1\right]}

\newcommand{\para}[1]{\left(#1\right)}

\newcommand{\Rg}{\ensuremath{R_{\rm g}}}

\newcommand{\nt}{\ensuremath{\tilde{N}}}

\newcommand{\pvn}{P_v(N)}

\newcommand{\fvn}{F_v(N)}

\newcommand{\fhyd}{F_{\rm hyd}}

\newcommand{\Nt}{\tilde{N}}
\newcommand{\Ntv}{\tilde{N}_v}

\newcommand{\kbt}{k_{\rm{B}} T}
\newcommand{\Rbar}{\overline{\textbf{R}}}

\begin{document}

\title{Characterizing Solvent Density Fluctuations in Dynamical Observation Volumes}

\author{Zhitong Jiang}
\affiliation{Department of Chemical and Biomolecular Engineering, University of Pennsylvania, Philadelphia, PA 19104, USA}
\author{Richard C. Remsing}
\affiliation{Department of Chemical and Biomolecular Engineering, University of Pennsylvania, Philadelphia, PA 19104, USA}
\affiliation{Institute for Computational Molecular Science, Temple University, Philadelphia, PA 19122, USA}
\author{Nicholas B. Rego} 
\affiliation{Biochemistry and Molecular Biophysics Graduate Group, University of Pennsylvania, Philadelphia, PA 19104, USA}
\author{Amish J. Patel}
\affiliation{Department of Chemical and Biomolecular Engineering, University of Pennsylvania, Philadelphia, PA 19104, USA}
\email[]{amish.patel@seas.upenn.edu}

\begin{abstract}
Hydrophobic effects drive diverse aqueous assemblies, such as micelle formation or protein folding,
wherein the solvent plays an important role.
Consequently, characterizing the free energetics of solvent density fluctuations can lead to important insights into these processes.
Although techniques such as the indirect umbrella sampling (INDUS) method~(Patel et al. \textit{J. Stat. Phys.} \textbf{2011}, \textit{145}, 265--275) can be used to characterize solvent fluctuations 
in static observation volumes of various sizes and shapes,
characterizing how the solvent mediates inherently dynamic processes, 
such as self-assembly or conformational change, remains a challenge.
In this work, we generalize the INDUS method to facilitate the enhanced sampling of solvent fluctuations 
in dynamical observation volumes, whose positions and shapes can evolve.
We illustrate the usefulness of this generalization by characterizing water density fluctuations in 
dynamic volumes pertaining to the hydration of flexible solutes,
the assembly of small hydrophobes, and conformational transitions in a model peptide.
We also use the method to probe the dynamics of hard spheres. 
\end{abstract}

\maketitle

\raggedbottom


\section{Introduction}
%
Solvent-mediated collective phenomena, such as hydrophobic effects, are important in a wide variety of contexts~\cite{dill_rev02,Chandler:Nature:2005,garde_rev,berne_rev09,Jamadagni:ARCB:2011,Pablo_Review} ranging from detergency~\cite{MaibaumDinnerChandler2004,meng2013pressure} and colloidal assembly~\cite{Rabani:2003,Morrone:2012}, to protein folding~\cite{Levy:2006aa,Dill:2012aa} and aggregation~\cite{shea08,Thirumalai:2012}.
Consequently, numerous theoretical, simulation, and experimental studies have been devoted to uncovering the molecular underpinnings of hydrophobic hydration and interactions~\cite{Li:PNAS:2011,davis2012water,ma2015modulation,tang2017guest}.
The hydration of hydrophobic solutes is unfavorable because they disrupt the favorable 
hydrogen bonding interactions between water molecules;
to minimize such disruption, hydrophobic moeities are also driven together to assemble in water.
Although hydrophobic solutes attract water through weak dispersive interactions,
they primarily disrupt water structure by excluding water molecules from the region they occupy~\cite{Remsing:JCP:2015}.
Consequently, purely repulsive hard solutes (HS) have long served as idealized hydrophobic solutes, 
and their hydration and assembly have been extensively studied using molecular simulations~\cite{Hummer:PNAS:1996,HGC,Patel:JPCB:2010,Remsing:2013,Patel:JPCB:2014} and theory~\cite{FHS:1973,Pratt:JCP:1977,ashbaugh_SPT,LCW}.

Because hydrating a hard solute is equivalent to emptying out a volume $v$ that it can occupy,
its hydration free energy, $\fhyd$, is related to the probability, $\pvn$, of observing $N$ waters in $v$ 
through~\cite{Widom:JCP:1963,Hummer:PNAS:1996}: $\beta\fhyd = -\ln P_v(N = 0)$, where $\beta = 1/k_{\rm B}T$, $k_{\rm B}$ is Boltzmann's constant, and $T$ is temperature. 
Thus, an understanding of the statistics of water density fluctuations, as quantified by $\pvn$,
has played a central role in elucidating the thermodynamic driving forces that underpin hydrophobic effects.
%
For example, Hummer et al.~\cite{Hummer:PNAS:1996, Garde:PRL:1996} showed that $\pvn$ is Gaussian
in small observation volumes ($v\lesssim1$~nm$^3$), enabling $\fhyd$ to be obtained from its moments. 
This powerful simplification  afforded molecular insights into diverse phenomena, 
including the formation of clathrate hydrates at high pressures~\cite{Hummer:PNAS:1998},
and the convergence of protein unfolding entropies at a particular temperature~\cite{Garde:PRL:1996,Remsing:2013}.
For larger volumes ($v\gtrsim1$~nm$^3$), both theory~\cite{LCW,LLCW,Geissler:2014:PRL,Geissler:2016:PNAS,Xi:PNAS:2016} 
and simulations~\cite{Patel:JPCB:2010,Patel:JSP:2011} 
have shown that the likelihood of low-$N$ fluctuations is enhanced significantly (relative to Gaussian) in bulk water, 
and even more so near hydrophobic surfaces; 
that is, although $\pvn$ remains Gaussian near its mean, it develops fat tails at low-$N$~\cite{Patel:JPCB:2010,Patel:JSP:2011}. 
These results clarified that water near hydrophobic surfaces sits at the edge of a dewetting transition, 
which can be triggered by small perturbations~\cite{Patel:JPCB:2012,berne04,berne05_melittin,chou_dewet}. 
They also led to the prediction that the assembly of small hydrophobic solutes in the vicinity of extended hydrophobic surfaces would be barrierless~\cite{Patel:PNAS:2011,Vembanur:2013}. 
Thus, an understanding of density fluctuations in small and large volumes as well as in the vicinity of interfaces has made significant contributions to our understanding of hydrophobic hydration, interactions and assembly.

%
Although $\pvn$ in small volumes can be readily estimated from unbiased simulations, 
the low-$N$ fluctuations of interest become increasingly rare for larger volumes, 
making it impossible to sample them without the use of non-Boltzmann or 
umbrella sampling methods~\cite{Jarzynski:2010:JPCB,Xi:JCTC:2016,Xi:Molsim:2018}.
However, biasing an order parameter such as the number of waters, $N_v$, 
in an observation volume, $v$, is complicated by the fact that 
$N_v$ is a discontinuous function of particle coordinates. 
Thus, choosing the biasing potential to be a function of $N_v$ would result in impulsive forces, which are difficult to treat in molecular dynamics (MD) simulations. 
The INDUS (INDirect Umbrella Sampling) method circumvents this challenge by sampling $N_v$ indirectly;
the biasing potential is chosen to be a function of $\nt_v$, which varies continuously with particle coordinates,
and is strongly correlated with $N_v$~\cite{Patel:JPCB:2010,Patel:JSP:2011}.
The particular manner in which $\nt_v$ is defined in INDUS, makes it amenable to Boolean algebra, so that the observation volume, $v$, can be constructed from unions and intersections of regular sub-volumes and their complements. 
This feature has facilitated the widespread use of INDUS for characterizing density fluctuations 
in volumes of various shapes and sizes in water~\cite{Patel:JPCB:2010,Patel:JSP:2011} and in other solvents~\cite{wu2014lengthscale}, at surfaces of various kinds, including electrodes~\cite{limmer2013hydration}, clay minerals~\cite{Rotenberg:JACS:2011}, and proteins~\cite{Patel:JPCB:2012,Remsing:JPCB:2018,yu2012simulations}, and in hydrophobic confinement~\cite{Remsing:PNAS:2015,Prakash:PNAS:2016}.

%
Although the observation volume used in INDUS, can have a wide variety of sizes and shapes,
it must be placed at a fixed location in the simulation box, and have a fixed, well-defined shape.
Such a static $v$ cannot be used to study the assembly of hydrophobic solutes,
or the hydration of flexible molecules, such as polymers or peptides, 
whose hydration shells change with their conformation.
To address this challenge, here we generalize INDUS to probe density fluctuations in dynamical observation volumes,
which can not only evolve with the positions of assembling molecules, but also conform continuously to the fluctuating shapes of flexible molecules.
%
%
We do so by pegging the position of $v$ (or the positions of the sub-volumes that make up $v$) to particles of interest.
When a biasing potential is applied, it exerts forces not only on the waters, but also on such particles;
as the particles move under the influence of the biasing force, the position (and shape) of $v$ also evolves.
We demonstrate the usefulness of this dynamic INDUS approach by characterizing density fluctuations 
in dynamical volumes in bulk water,
in the hydration shells of small hydrophobic solutes undergoing assembly,
and in the hydration shell of the conformationally flexible alanine dipeptide molecule.
We also illustrate the use of this framework to mimic hard sphere systems and study their dynamics.

\section{Generalizing INDUS to Dynamic Observation Volumes}
%
Although the methods discussed here are applicable to any order parameter, 
we focus on the number of waters, $N_v$, in an observation volume, $v$, of interest.
We wish to estimate the free energetics, $\beta\fvn = -\ln \pvn$, where 
$\pvn = \langle \delta(N_v - N) \rangle_0$, is the probability of observing $N$ waters in $v$.
Given the Hamiltonian of the system, $\mathcal{H}_0(\Rbar)$, where $\Rbar$ represents the positions of all atoms,
$\langle \mathcal{O}(\Rbar) \rangle_0 \equiv \int d\Rbar~\mathcal{O}(\Rbar) \exp(-\beta \mathcal{H}_0) / Q_0$ is the ensemble average of $\mathcal{O}(\Rbar)$, and $Q_0\equiv \int d\Rbar~\exp(-\beta \mathcal{H}_0)$ is the system partition function.

\subsection{Umbrella Sampling vs Indirect Umbrella Samling}
%
Umbrella sampling and similar non-Boltzmann sampling techniques employ biasing potentials, 
$U(N_v)$, which enhance the likelihood with which rare order parameter fluctuations are sampled~\cite{Jarzynski:2010:JPCB}. 
To apply biasing potentials in molecular dynamics (MD) simulations, 
the order parameter being biased must be a continuous and differentiable function of particle positions.
However, $N_v$ is a discrete function of particle positions, as can be seen from:
\begin{equation}
N_v(\{ \rb_i \}) = \sum_{i=1}^M h_v(\rb_i),
\end{equation}
where $M$ is the total number of waters in the system, $\rb_i$ is the position of the $i^{\rm th}$ water, 
and $h_v(\rb_i)$ is an indicator function which is equal to 1 if $\rb_i \in v$ and 0 otherwise.
Because $h_v$ (and thereby $N_v$) changes discontinuously as water molecules cross the boundary of $v$, 
biasing $N_v$ would result in impulsive forces on such waters.

To circumvent this challenge, Indirect Umbrella Sampling (INDUS)~\cite{Patel:JPCB:2010,Patel:JSP:2011} prescribes 
influencing $N_v$ indirectly by biasing a closely related continuous variable $\Ntv$, which is defined as
\begin{equation}
\Ntv(\{ \rb_i \}) = \sum_{i=1}^M \tilde{h}_v(\rb_i).
\end{equation}
Here, $\tilde{h}_v(\rb_i)$ is a coarse-grained (or smoothed) indicator function, which varies continuously with the particle positions, $\{\rb_i\}$. 
Akin to the discrete indicator function, $h_v(\rb_i)$, the continuous indicator function $\tilde{h}_v(\rb_i)$ is 1 when $\rb_i$ is well within $v$, and 0 when it is well removed from $v$.
However, in contrast with $h_v(\rb_i)$, the smoothed indicator function, $\tilde{h}_v(\rb_i)$, 
changes continuously from 1 to 0 as particle $i$ leaves $v$.

The INDUS method prescribes factoring the coarse-grained indicator function $\tilde{h}_v(\rb_i)$ 
into independent contributions from spatial coordinates as
\begin{equation}
\tilde{h}_v(\rb_i) = \prod_{\alpha} \tilde{h}_v^{(\alpha)}(\alpha_i).
\end{equation}
Here, $\tilde{h}_v^{(\alpha)}(\alpha_i)$ represents a coordinate indicator function, 
with the coordinates chosen considering the shape of $v$; 
e.g., Cartesian coordinates are recommended for a cuboidal $v$, with $\alpha$ representing $x$, $y$, and $z$.
The coordinate indicator functions  then report whether $\rb_i \in v$ along the respective coordinate; e.g., if $x_i$ is well within the $x$-bounds that define a cuboidal $v$, $\tilde{h}_v^{(x)}(x_i)$ would be 1.
Note that factors for each of the three spatial dimensions are not always necessary.
In particular, for a spherical $v$, only the radial coordinate, $r$, is needed.
In this case, $\tilde{h}_v^{(\theta)}$ and $\tilde{h}_v^{(\phi)}$ are set to 1, 
and $\tilde{h}_v(\rb_i) = \tilde{h}_v^{(r)}(r_{i{\rm C}})$, where $r_{i{\rm C}} = |\rb_i - \rb_{\rm C}|$, and $\rb_{\rm C}$ denotes the center of $v$.

In principle, any functional form that increases continuously from 0 to 1 as the particle enters $v$, 
and decreases continuously from 1 to 0 as it leaves $v$, can be used to represent $\tilde{h}_v^{(\alpha)}(\alpha_i)$.
In ref.~\cite{Patel:JSP:2011}, $\tilde{h}_v^{(\alpha)}(\alpha_i)$ was defined as the integral
of a Gaussian coarse-graining function with a standard deviation of $\sigma$, which was truncated and shifted to zero at $\alpha_{\rm c}$, and then normalized; 
for this choice, analytic expressions for $\tilde{h}_v^{(\alpha)}(\alpha_i)$ 
and their derivatives can also be found in ref.~\cite{Patel:JSP:2011}.
As $\sigma\rightarrow 0$, $\tilde{h}_v(\rb_i) \to h_v(\rb_i)$, and $\Ntv \to N_v$.
Thus, a judicious choice of the parameter $\sigma$ is one that is sufficiently small that $\Ntv$ closely follows $N_v$, 
but not so small that biasing $\Ntv$ gives rise to large forces that cannot be handled in MD simulations.
This balance is typically achieved by choosing $\sigma$ to be a fraction of the particle radius, and $\alpha_{\rm c}$ to be 2 to 3 times $\sigma$.
%

\subsection{Static vs Dynamic INDUS}
%
The order parameter, $\Ntv$, which is biased using INDUS, represents the coarse-grained particle number
in an observation volume that is static; that is, $v$ is placed at a fixed location in the simulation box.
A biasing potential, $U(\Ntv)$, typically chosen to be harmonic or linear in $\Ntv$, 
is then employed to sample $\Ntv$ (and thereby $N_v$) over its entire range of interest.
The force, $\mathbf{f}_i$, on the $i^{\rm th}$ water arising from $U(\Ntv)$ is given by:
\begin{equation}
\mathbf{f}_i = -\nabla_i U = -\frac{\partial U}{\partial \Ntv} \cdot \nabla_i \tilde{h}_v(\rb_i),
\end{equation}
where $\nabla_i$ represents the gradient with respect to $\rb_i$, the position of particle $i$.

For a spherical $v$, $\tilde{h}_v(\rb_i) = \tilde{h}_v^{(r)}(r_{i{\rm C}})$, so the expression for the force simplifies to:
\begin{equation}
\mathbf{f}_i = -\frac{\partial U}{\partial \Ntv} \cdot \frac{ \partial \tilde{h}_v^{(r)} }{ \partial r_{i{\rm C}} } \cdot \frac{\rb_{i{\rm C}}}{r_{i{\rm C}}},
\end{equation}
where $\rb_{i{\rm C}} \equiv \rb_i - \rb_{\rm C}$, $r_{i{\rm C}} = |\rb_{i{\rm C}}|$, and $\rb_{\rm C}$ is the center of the spherical $v$.

%
We now generalize INDUS to characterize density fluctuations in dynamical observation volumes, which can evolve, 
e.g., with the positions of assembling molecules.
To do so, we peg the position of $v$ to a mobile particle.
In particular, for a spherical $v$, we peg the center of the sphere to a particle,
which can either be an ideal (dummy) particle or a real (interacting) particle.
In other words, the center of the spherical observation volume, $\rb_{\rm C}$,
is now associated with the position of a particle.
When a biasing potential is applied, it exerts forces not only on the waters, 
but also on the particle that is pegged to the center of $v$.
In fact, it can readily be seen that the biasing force on that particle, $\mathbf{f}_{\rm C} \equiv - \nabla_{\rm C} U$, is equal and opposite to the sum of the forces on all the waters, and is given by
\begin{equation}
\mathbf{f}_{\rm C} = -\sum_{i=1}^M \mathbf{f}_i.
\end{equation}
By using $\mathbf{f}_{\rm C}$ to evolve the position, $\rb_{\rm C}$, of the center of $v$, under the equations of motion used in the simulation, 
we can then sample the fluctuations of $\Ntv$ (and thereby $N_v$) in dynamic observation volumes.

The above discussion focussed on a spherical observation volume; however the underlying principles are general,
and can be used to extend INDUS to dynamical observation volumes of other regular shapes, such as cuboids or cylinders.
Although we do not undertake this exercise here,
below we demonstrate the extension of these ideas to an irregularly-shaped $v$ that is comprised of a union of spherical sub-volumes.
Thus, INDUS can also be generalized to dynamical observation volumes obtained by performing Boolean operations on regularly-shaped sub-volumes.

\subsection{Dynamic INDUS for Union of Spheres}
%
To construct a dynamical observation volume, $v$, using the union of $S$ spherical sub-volumes,
we peg the centers of each of the $S$ sub-volumes to the positions, $\rb_p$, of $S$ mobile particles ($p = 1, 2, \dots, S$).
Following ref.~\cite{Patel:JSP:2011}, the coarse-grained indicator function is given by
\begin{equation}
\tilde{h}_v(\rb_i) = 1 - \prod_{p=1}^S [1 - \tilde{h}_p(\rb_i) ].
\end{equation}
Here, $\tilde{h}_p(\rb_i)$ is the coarse-grained indicator function for the $p^{\rm th}$ sub-volume, and for spherical sub-volumes, it simplifies to
\begin{equation}
\tilde{h}_p(\rb_i) = \tilde{h}^{(r)}_p(r_{ip}),
\end{equation}
with $r_{ip} \equiv | \rb_i - \rb_p |$. 
Then, by recognizing that $\Ntv = \sum_{i=1}^M \tilde{h}_v(\rb_i)$,
the biasing force on the $i^{\rm th}$ water arising from a potential, $U(\Ntv)$, can be obtained as

\begin{equation}
\mathbf{f}_i = -\frac{\partial U}{\partial \Ntv} \cdot \bigg[ \sum_{p=1}^S \prod_{q \ne p} [1 - \tilde{h}_q(\rb_i) ] \cdot \frac{ \partial \tilde{h}_p^{(r)} }{ \partial r_{ip} } \cdot \frac{\rb_{ip}}{r_{ip}} \bigg].
\end{equation}
We note that $\mathbf{f}_i$ can be expressed as a sum of contributions from each of the $S$ sub-volumes,
\begin{equation}
\mathbf{f}_i = \sum_{p=1}^S \mathbf{f}_{ip},
\end{equation}
where the contribution of the $p^{\rm th}$ sub-volume to the total force on the $i^{\rm th}$ water, $\mathbf{f}_{ip}$, is given by:

\begin{equation}
\mathbf{f}_{ip} = -\frac{\partial U}{\partial \Ntv} \cdot \prod_{q \ne p} [1 - \tilde{h}_q(\rb_i) ] \cdot \frac{ \partial \tilde{h}_p^{(r)} }{ \partial r_{ip} } \cdot \frac{\rb_{ip}}{r_{ip}}.
\end{equation}
The forces on the particles pegged to centers of sub-volumes can then be readily obtained as:
\begin{equation}
\mathbf{f}_{p} = -\sum_{i=1}^M \mathbf{f}_{ip},
\end{equation}
i.e., the force on the particle pegged to the $p^{\rm th}$ sub-volume is equal and opposite 
to the sum of the forces exerted by that sub-volume on all the waters.

\section{Methods}
\label{sec:methods}
All-atom MD simulations are performed using the GROMACS package (version 4.5.3)~\cite{gmx4ref}, 
suitably modified to incorporate the biasing forces derived in the preceding section. 
The leap frog algorithm~\cite{Frenkel_Smit} was used to integrate the equations of motion with a 2 fs time-step;
the motion of the center of mass of the system was removed every 10 time-steps, 
and periodic boundary conditions were employed in all dimensions.
The SPC/E model of water~\cite{SPCE} was employed in all simulations, with the water molecules being constrained to be rigid using the SHAKE algorithm~\cite{SHAKE}.
Long range electrostatic interactions were computed using the Particle Mesh Ewald algorithm~\cite{PME},
and short range Lennard-Jones and electrostatic interactions were truncated at 1~nm.
Lorentz-Berthelot mixing rules were used to model the cross-interactions between solutes and SPC/E water. 
Simulations of alanine dipeptide were performed in the NVT ensemble; whereas all other simulations were performed in the NPT ensemble.
The system temperature was maintained at 300~K using the canonical velocity-rescaling thermostat in all cases~\cite{Bussi:JCP:2007}.
For the NPT simulations, a pressure of 1~bar was maintained using the Parrinello-Rahman barostat~\cite{Parrinello-Rahman}.
We now describe details specific to the various systems that we study.

{\bf Spherical Solute:}
To begin with, a spherical observation volume with radius, $R_v=0.3$~nm, was placed at the center of a cubic simulation box with a side length of 4~nm. 
In the dynamic INDUS simulations, the center of $v$ was pegged to a dummy atom, and was thereby mobile, whereas in the static INDUS simulations, the center of $v$ remained at its initial position.
The dummy atom was chosen to have a mass of 28.054. 
To allow for equilibration, the first 0.1~ns were discarded from each simulation.
%

{\bf Hard-Sphere Alkane:}
The united-atom n-hexadecane chain was modeled using the TraPPE-UA (Transferable Potentials for Phase Equilibria - United Atom) force field~\cite{trappe-ua}.
A spherical sub-volume of radius, $R_v=0.4$~nm, was centered on each of the 16 united-atoms,
and the union of the spherical sub-volumes served as the dynamical observation volume, $v$,
whose size and shape evolved with the conformation of the alkane.
The interactions between the alkane and SPC/E water were turned off. 
Dynamic INDUS simulations of the alkane in water were then performed to obtain the free energetics of $N_v$-fluctuations;
each biased simulation was run for 5~ns with the first 0.5~ns being discarded to allow for equilibration.
To independently characterize the conformational free energy landscape of the alkane, 
a 10~ns long unbiased simulation of the alkane in vacuum was performed in the NVE ensemble. 
In both cases, a cubic simulation box with a side length of 6~nm was used.
%
%

{\bf Assembly of Small Hydrophobic Solutes:}
The assembly of methanes as the prototypical small solutes was studied in bulk water.
Methanes are simulated as united atoms using the TraPPE-UA force field~\cite{trappe-ua}.
We also study purely repulsive methanes, which interact using only the repulsive portion of the WCA potential~\cite{wca}.
The simulation box is 3~nm in each direction, and contains 13 methane (attractive or repulsive) molecules and 868 SPC/E waters.
Spherical sub-volumes of radius, $R_v=0.5$~nm, were centered on each of the 13 solutes,
and the union of the spherical sub-volumes served as the dynamical observation volume, $v$.
Dynamic INDUS simulations were then performed to obtain the free energetics of $N_v$-fluctuations;
each biased simulation was run for 2~ns with the first 0.5~ns being discarded to allow for equilibration.

{\bf Alanine Dipeptide:}
The AMBER99SB force field was used to simulate alanine dipeptide~\cite{AMBER,Lange_amber99sb}.
A roughly $3$~nm cubic water slab containing 882~waters was used to solvate the peptide.
Simulations were performed using a $3 \times 3 \times 6$~nm$^3$ simulation box 
with a repulsive wall at its $z$-edge to nucleate buffering vapor layers.
The presence of vapor-liquid interfaces keeps the system at its coexistence pressure.
The centers-of-masses of alanine dipeptide and the water slab were restrained in the $z$-direction to be at the center of the simulation box.
The observation volume $v$ was defined as a union of spherical sub-volumes of radius $0.6$ nm, centered on the peptide's 10 heavy atoms.
The hydration of alanine dipeptide was modulated by applying a linear biasing potential, $\alpha \tilde{N}_v$.
A total of $15$ different biasing strengths ranging from $\beta\alpha=0$ to $8$ were used.
At each value of $\alpha$, the conformational degrees of freedom of alanine dipeptide, as represented by the backbone dihedral torsional angles, $\Phi$ and $\Psi$, were sampled using standard umbrella sampling with harmonic umbrella potentials.
Roughly 96 biased simulations were run for each $\alpha$; the simulations were run for $5$~ns, with the first $0.5$~ns excluded from analysis.
The conformational free energy landscape was obtained from the biased simulations using unbinned WHAM (UWHAM)~\cite{UWHAM} or equivalently, the multi-state Bennett Acceptance Ratio (MBAR) method~\cite{MBAR}.

\begin{figure*}[htb]
\centering
\includegraphics[width=0.95\textwidth]{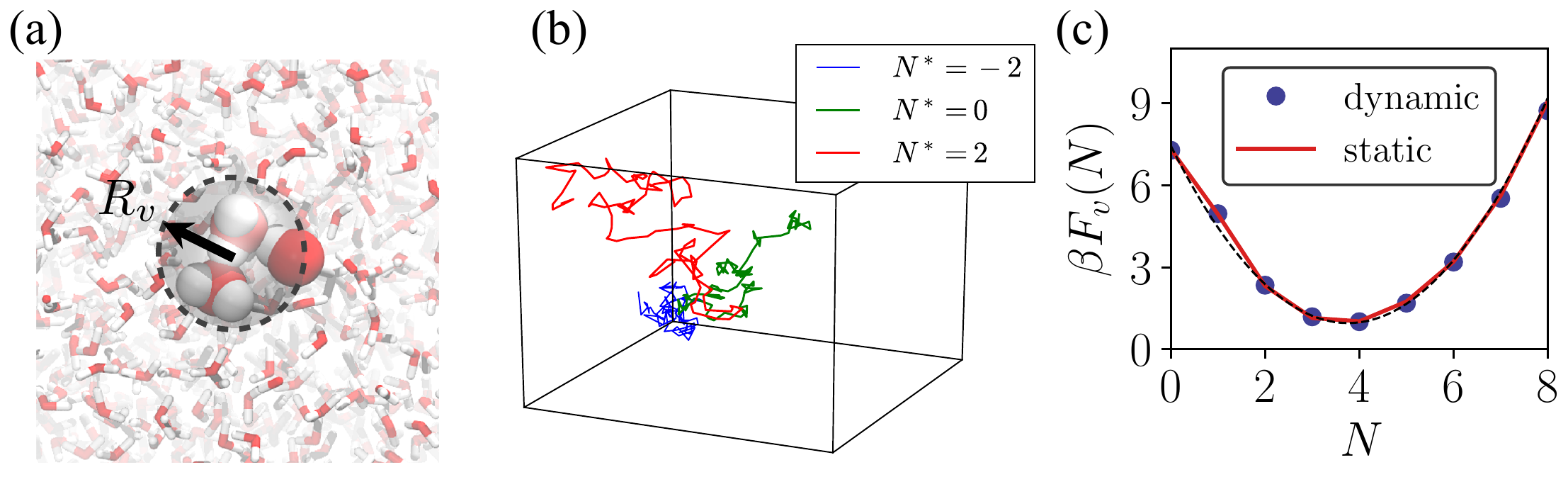} 
\caption{
(a) Simulation snapshot of a small spherical observation volume, $v$, of radius, $R_v = 0.3$~nm, in bulk water. 
The center of $v$ is pegged to the position of a dummy atom.
The waters inside $v$ are shown in the space-fill representation, whereas those outside $v$ are shown in the licorice representation. 
(b) When a harmonic biasing potential, $U_{\kappa,N^*}$, is applied to waters in the dynamical observation volume, 
the waters in $v$ experience a biasing force, but so does the dummy atom pegged to the center of $v$.
Trajectories depicting the motion of the dummy atom are shown here for three different biasing potentials.
(c) The free energy, $\beta F_v(N) \equiv -\ln P_v(N)$, obtained using INDUS with a dynamical $v$, 
agrees well that obtained using a static $v$;
moreover, $\fvn$ is parabolic (dashed line), highlighting that $\pvn$ for this small volume is Gaussian.
}
\label{fig:hydr}
\end{figure*}

\section{Hydrophobic Hydration: Fluctuations in Bulk Water}

The free energetics of water density fluctuations, $\fvn$, in bulk water inform hydrophobic hydration.
In particular, the free energetic cost of emptying $v$, 
corresponds to the hydration free energy, $F_{\rm hyd}$,
of a purely repulsive hydrophobic solute, which has the same size and shape as $v$,
i.e., $F_{\rm hyd} = F_v(N\to0)$.
Such idealized hydrophobic solutes, which simply exclude waters from the region they occupy,
have long been used to study hydrophobic hydration and interactions because the extent to which
they disrupt the proximal water structure is similar to that of real hydrophobic solutes~\cite{Remsing:JCP:2015}.
In this section, we characterize $\fvn$ in two observation volumes in bulk water,
which correspond to a small hard sphere and a repulsive n-hexadecane molecule as $N\to0$.

\subsection{Ideal Spherical Solute}
%
To illustrate the dynamic INDUS method, we first use it to estimate $\fvn$ in a small spherical volume in bulk water (Figure~\ref{fig:hydr}a).
To estimate $\fvn$, we employ a dynamical observation volume of radius, $R_v = 0.3$~nm, with its center pegged to a mobile dummy (ideal gas) atom.
When a biasing potential, $U_{\kappa,N^*}(\Ntv) = \frac{\kappa}{2} (\Ntv - N^*)^2$, is applied, 
it gives rise to forces on the water molecules in $v$, 
but also on the dummy atom. 
Trajectories of the dummy atom for $\beta \kappa = 1.12$ and select $N^*$-values are shown in Figure~\ref{fig:hydr}b; 
interestingly, as $N^*$ is decreased, and $v$ is emptied, its mobility appears to decrease as well.
Following the INDUS prescription, biased simulations are then used to obtain $\fvn$ for this dynamical observation volume.
As shown in Figure~\ref{fig:hydr}c, the fluctuations display Gaussian statistics in agreement with previous studies~\cite{Hummer:PNAS:1996,Patel:JSP:2011}.
The hydration free energy of the cavity that forms as $N_v \to 0$ (and of the corresponding hard sphere solute) 
is $F_{\rm hyd} = 7.3~\kbt$.
Also shown in Figure~\ref{fig:hydr}c, are the free energetics obtained using a static observation volume, which is pegged to the center of the simulation box.
As expected from the translational invariance of observation volumes in bulk water, 
we find that the $\fvn$-profiles obtained using the dynamic and static volumes are in excellent agreement with one another.

\begin{figure*}[htb]
\centering
\includegraphics[width=0.99\textwidth]{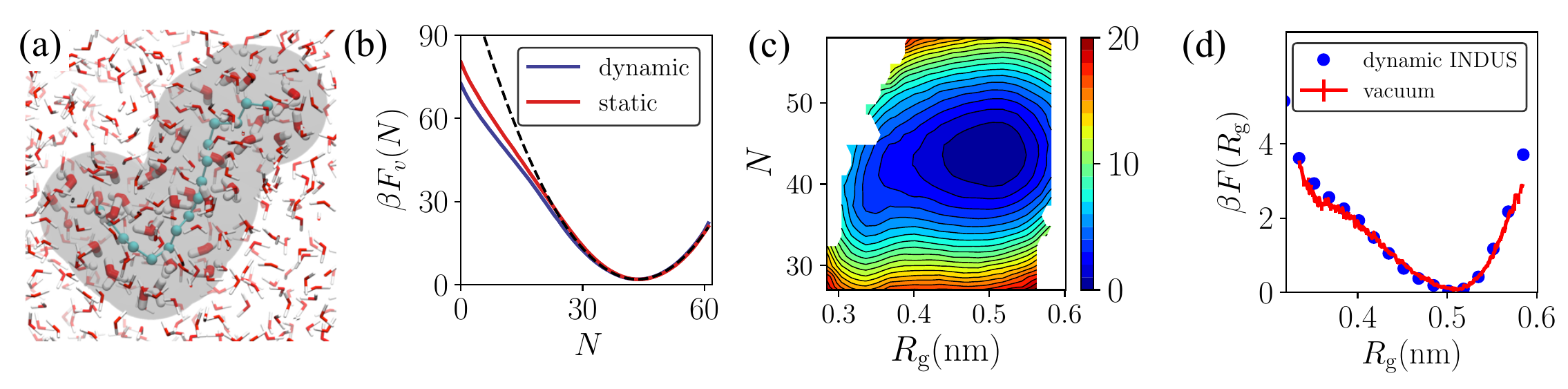} 
\caption{
(a) Simulation snapshot illustrating a dynamical observation volume, $v$ (gray), in bulk water.
To define $v$, we use a dummy united-atom n-hexadecane molecule (cyan);
spherical sub-volumes of radius, $R_v = 0.4$~nm, are centered on each of the united-atoms of the alkane, 
and $v$ is defined as the union of the 16 sub-volumes.
Water molecules (red/white)
inside $v$ are shown in licorice representation, 
whereas those outside $v$ are shown as lines. 
(b) The free energetics $F_v(N)$ of water number fluctuations in bulk water, obtained using this dynamical (flexible) $v$, are shown,
and are compared to those obtained using a static (rigid) observation volume.
In both cases, the statistics of water number fluctuations are Gaussian (i.e., $\fvn$ are parabolic, dashed line) near the mean of the distributions, 
but display fat tails for the lower $N$-values; the low-$N$ tail is fatter for the dynamical $v$.
(c) Free energy, $F_v(N,\Rg)$, is shown as a function of the number of waters, $N$, in $v$, 
and the the radius of gyration, $\Rg$, of the alkane;
contours are drawn in increments of $1~\kbt$.
As $N$ decreases, the alkane adopts configurations with smaller $\Rg$-values.
(d) The free energy $F(\Rg)$ obtained by integrating $\exp[-\beta F_v(N,\Rg)]$ over the solvent co-ordinate, $N$,
agrees well the corresponding $F(\Rg)$ obtained by simulating the alkane in vacuum.
}
\label{fig:alkane}
\end{figure*}

\subsection{Hard-Sphere Alkane}
Next, we characterize water density fluctuations in an observation volume whose shape and size can vary. 
The dynamical $v$ is defined as the union of 16 spherical sub-volumes; each sub-volume is pegged to a dummy united-atom of n-hexadecane, and has a radius, $R_v = 0.4$~nm (Figure~\ref{fig:alkane}a).
The dummy alkane chain has the same intra-molecular interactions (bonded and non-bonded) 
as a united-atom alkane~\cite{trappe-ua}, but its atoms do not interact with water; 
thus, the alkane conformational free energy landscape, which governs the size and shape of $v$, 
is determined entirely by the self-interactions of the alkane.
We estimate the free energetics, $\fvn$, of water number fluctuations in this $v$ in bulk water using dynamic INDUS; see Figure~\ref{fig:alkane}b.
On average, $v$ contains roughly 44 waters, and although water density fluctuations are Gaussian 
near the mean and at high $N$, they are enhanced at low $N$ and display marked fat tails, 
suggesting that the dewetting of this dynamical $v$ is co-operative~\cite{Patel:JPCB:2012}. 
Although our choice of $R_v = 0.4$~nm is somewhat larger than (but close to) the hard-core radii of the alkane united-atoms,
we obtain the hydration free energy of this conformationally flexible hard-sphere n-hexadecane molecule to be $\beta F_{\rm hyd}=70.4$.
%

To uncover the extent to which the collective dewetting of $v$ is facilitated by its conformational degrees of freedom, 
we additionally estimate $\fvn$ for a static observation volume defined using a representative n-hexadecane configuration.
As shown in Figure~\ref{fig:alkane}b, although the low-$N$ fluctuations for the static $v$ 
are also enhanced relative to Gaussian statistics, they are suppressed relative to the dynamic $v$.
The difference in estimates of $F_v(N\to0)$ for the static and dynamic volumes ($78.2~\kbt$ vs $70.4~\kbt$), 
represents the difference in the hydration free energies of rigid and flexible hard-sphere alkanes ($7.8~\kbt$).
We thus find the solvation of the flexible alkane to be more favorable than that of the rigid alkane.
Our findings are consistent with those of Pettitt and co-workers, who showed that the conformation of alkane chains
and peptides can influence their hydration free energies substantially~\cite{Pettitt:PNAS,Kokubo:JPCB:2013,Asthagiri_2017}.
Our results also lend further support to the notion that surface-area models, 
which are commonly used to estimate the driving force of hydrophobic assembly,
but are incapable of capturing subtle but important differences between the rigid and flexible solutes,
are not appropriate for a quantitative treatment of hydrophobic hydration and interactions~\cite{Pettitt:PNAS,Xi:PNAS:2017}.
%

To further understand how the dewetting of $v$ influences the configuration of the flexible alkane (characterized using its radius of gyration, $\Rg$), we the use the dynamic INDUS simulations to estimate free energy as a function of $N$ and $\Rg$ (Figure~\ref{fig:alkane}c).
The free energetics display a well-defined basin at $N = 44$ and $\Rg = 0.5$~nm.
At the highest values of $N$, when $v$ is filled with waters, the alkane is extended, and displays a high value of $\Rg$.
In contrast, at low values of $N$, the alkane collapses onto itself as much as possible to minimize the cost of emptying $v$.
With the alkane collapsed to $\Rg = 0.32$~nm, roughly 30 waters still remain in $v$.
These remaining waters are then displaced with little change in $\Rg$ to bring about the complete drying of $v$.
Because the system has no alkane-water interactions, 
integrating out the solvent degrees of freedom (represented by $N$) 
ought to give rise to the same configurational landscape, $F(\Rg)$, 
that would be obtained by simulating the alkane in vacuum;
this is indeed the case, as illustrated by the excellent agreement between the two quantities (Figure~\ref{fig:alkane}d).
%

\begin{figure*}[htb]
\centering
\includegraphics[width=0.99\textwidth]{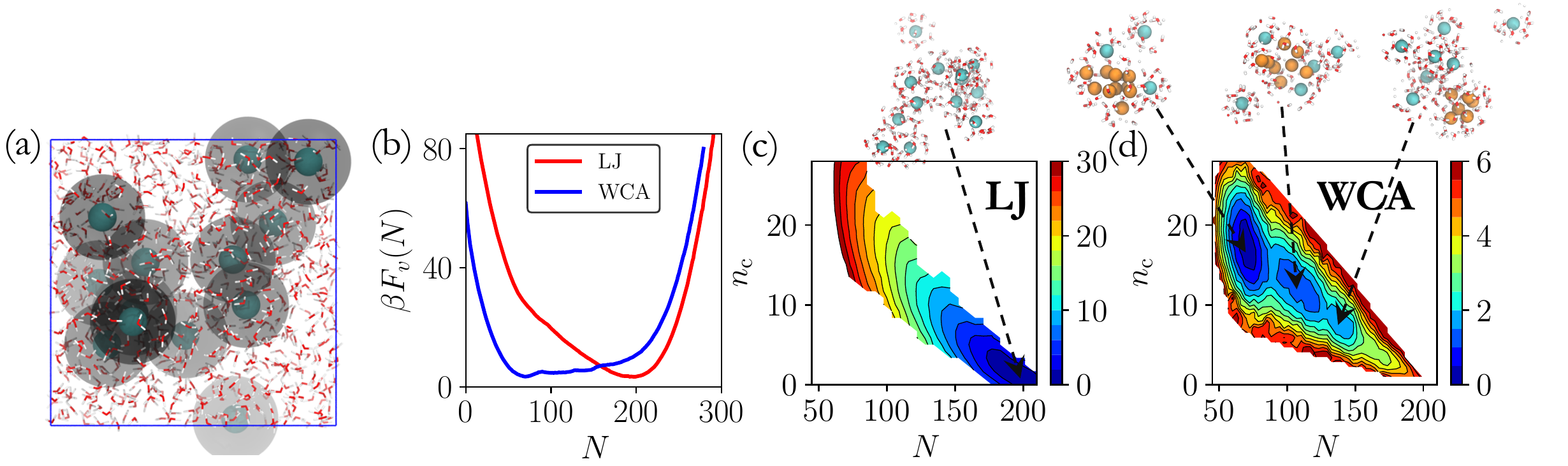} 
\caption{
(a) Simulation snapshot of the system shows 13 united-atom methanes (cyan) solvated in water (red/white lines). 
The observation volume $v$ is defined as the union of spherical sub-volumes (gray) of radii, $R_v = 0.5$~nm, that are centered on each of the methanes.
(b) The free energy, $\fvn$, is shown for both attractive (LJ) and purely repulsive (WCA) methanes.
The LJ solutes are well-hydrated, whereas the WCA solutes are not.
(c, d) The free energy landscape as a function of the number, $N$, of waters in $v$, and the number, $n_{\rm c}$, of solute-solute contacts is shown for the LJ~(c) and WCA~(d) methanes.
Methane molecules that participate in contacts are shown in orange, whereas hydrated methanes are colored cyan. 
Only the water molecules in $v$, i.e., waters within 0.5~nm of any methane molecules, 
are shown (in red/white); the remainder of the waters are hidden.
For both the LJ and WCA solutes, an increase in $N$ is correlated with a corresponding decrease in $n_{\rm c}$ for $80 < N < 200$ and $0 < n_{\rm c} < 15$;
the LJ solutes have a preference for being hydrated and dispersed, whereas the WCA solutes prefer to dewet and cluster.
}
\label{fig:methane1}
\end{figure*}

\section{Hydrophobic Assembly}
%
Hydrophobic solutes solvated in water disrupt the favorable hydrogen bonding interactions between water molecules.
To minimize this disruption of favorable water-water interactions, hydrophobic solutes tend to self-assemble,
forming the basis for diverse processes, ranging from supramolecular chemistry to protein folding~\cite{tang2017guest,Dill:2012aa}.
Here we use dynamic INDUS to elucidate the role of water density fluctuations in the assembly of small, hydrophobic solutes.
In contrast with the previous section, where we employed dummy atoms to define $v$, and characterized $\fvn$ in bulk water, 
here we will make use of real hydrophobic solutes to define $v$, and characterize $\fvn$ in their hydration shells.

\subsection{Role of Solute-Water Attractions}
%
Idealized, purely repulsive solutes, which simply exclude water, have long been used to study hydrophobic effects.
Such solutes themselves do not attract each other, but are driven by the solvent to aggregate in aqueous solution~\cite{Pratt:JCP:1977,ashbaugh_SPT,Chaudhari_2013,Gao:2018}.
In addition to excluding water, real non-polar solutes also possess favorable dispersive interactions with water.
Such solute-water attractive interactions offset the penalty associated with the disruption of water structure somewhat, 
and lower the driving force for hydrophobic assembly by stabilizing dispersed states.
Indeed, the driving force for the aqueous association of two small non-polar solutes (e.g., methanes) tends to be smaller than the thermal energy.

To elucidate the role of water in the assembly of small hydrophobic solutes, 
and uncover how the driving force for assembly is influenced by solute-water interactions,
here we study the assembly of 13 methane molecules with and without attractive interactions.
Spherical sub-volumes of radii, $R_v = 0.5$~nm are pegged to each of the methanes,
and the union of these sub-volumes is chosen to be our dynamical observation volume, $v$ (Figure~\ref{fig:methane1}a).
The observation volume thus represents the collective hydration shell of all the methanes.
In Figure~\ref{fig:methane1}b, the free energetics of water number fluctuations in $v$ 
are shown for both the attractive  (LJ) and purely repulsive (WCA) solutes.
We find that the attractive solutes tend to be well-hydrated, with a total of nearly 200 waters being observed in their hydration shells on average.
In contrast, for the repulsive solutes only about 70 waters populate $v$ on average.
%

To characterize the configurations that give rise to this behavior, we estimate the joint free energy, $F(N,n_{\rm c})$, 
where $n_{\rm c}$ is the number of solute-solute contacts, which are said to have formed when 
the centers of two solutes are separated by less than 0.45 nm.
As shown in Figure~\ref{fig:methane1}c, in their lowest free energy configurations, the attractive solutes are well-hydrated, and make few solute-solute contacts.
The loss of waters from $v$ is accompanied by the formation of solute-solute contacts, 
with the decrease in $N$ being correlated with the increase in $n_{\rm c}$ over $80 < N < 200$ and $0 < n_{\rm c} < 15$.
In contrast, the stable basin in $F(N,n_{\rm c})$ for the repulsive solutes corresponds to small $N$ and large $n_{\rm c}$ (Figure~\ref{fig:methane1}d),
i.e., the repulsive solutes prefer to cluster with the corresponding $v$ being relatively devoid of waters.
Nevertheless, as with the attractive solutes, $N$ and $n_{\rm c}$ are correlated over 
$80 < N < 200$ and $0 < n_{\rm c} < 15$ for the repulsive solutes.
Interestingly, a number of distinct metastable basins are also seen in Figure~\ref{fig:methane1}d for the repulsive solutes highlighting the presence of desolvation barriers that are characteristic of hydrophobic assembly.
%

For both the LJ and WCA methanes, once roughly 15 contacts are formed between the solutes, a further decrease in $N$ (below 80) does not lead to an increase in $n_{\rm c}$; instead the sharp increase in $\fvn$ upon decreasing $N$ below 80 waters (Figure~\ref{fig:methane1}b) corresponds to the dewetting of the cluster.
The sharp increase in $\fvn$ seen when $N$ is increased above 200 waters, on the other hand, 
corresponds to excess waters being squeezed into the hydration shells of individual solutes.
%
In the intermediate region from roughly 80 to 200 waters,
$\fvn$ captures the free energetics of the correlated aggregation and dewetting 
(or equivalently, dissociation and hydration) of the small, hydrophobic solutes.
%
Although the aggregation of WCA solutes is favorable ($\Delta F_{\rm agg} \approx -7~\kbt$) 
and that of the LJ solutes is not ($\Delta F_{\rm agg} \approx 20~\kbt$),
in both cases, these overall preferences arise from a competition between the
strong inherent tendency for the hydrophobic solutes to cluster in water (favorable),
and a substantial loss of mixing entropy upon aggregation (unfavorable).
For our simulations, which have 13 solutes in 868 waters, the ideal solution contribution to the 
loss of mixing entropy is roughly $60~\kbt$.
The inherent preference for clustering can then be quantified using  the excess aggregation free energy, 
which is $\Delta F_{\rm agg}^{\rm ex} \approx -67~\kbt$ for the WCA solutes 
and $\Delta F_{\rm agg}^{\rm ex} \approx -40~\kbt$ for the LJ solutes.

\begin{figure*}[htb]
\centering
\includegraphics[width=0.85\textwidth]{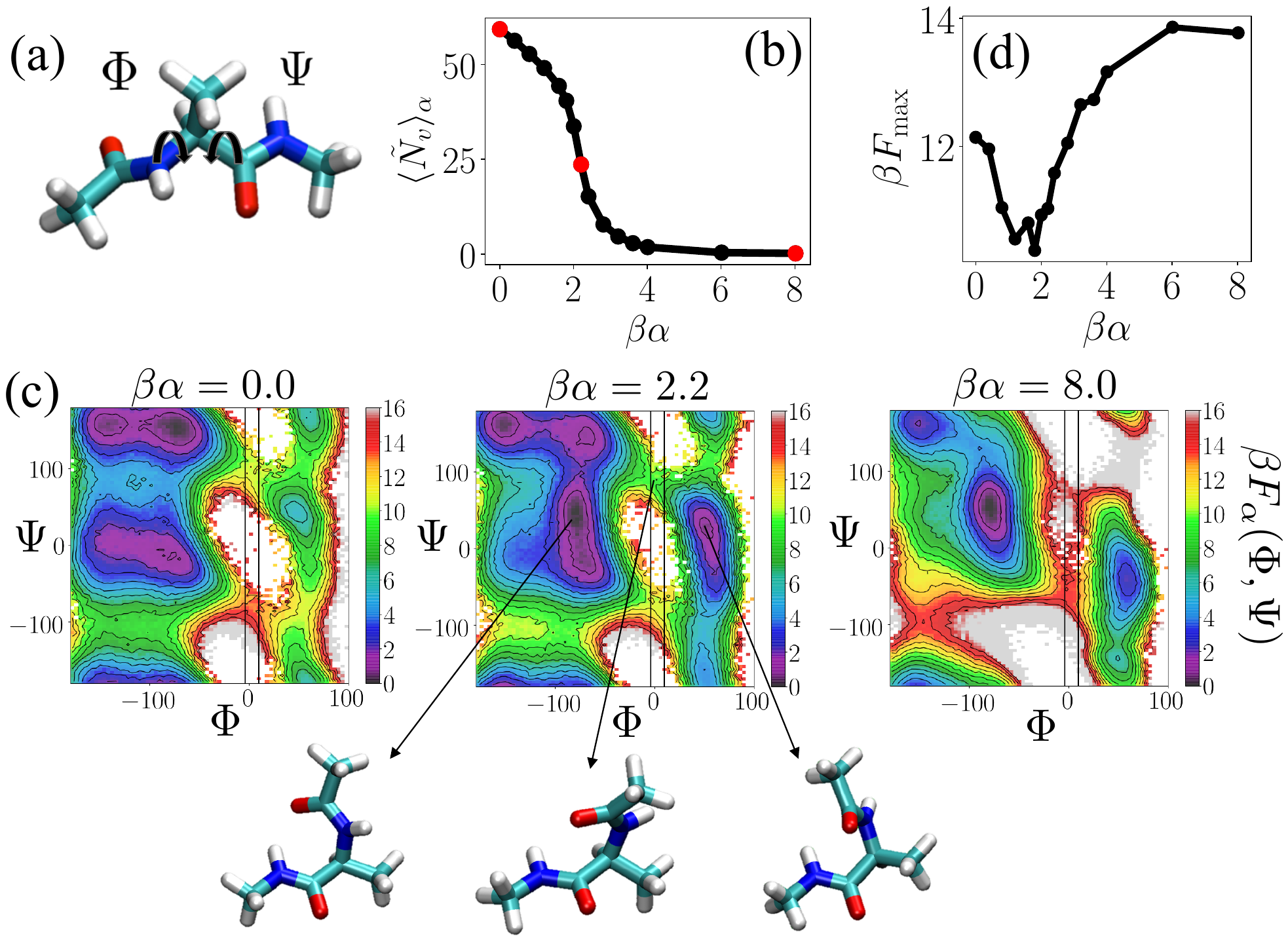} 
\caption{
(a) Alanine dipeptide is shown in licorice representation along with its two principal dihedral angles, $\Phi$ and $\Psi$;
the color scheme is: cyan = C, white = H, red = O, blue = N. 
(b) The response of the average number of waters, $\langle \tilde{N}_v \rangle_\alpha$, in the peptide hydration shell, $v$, to an unfavorable biasing potential, $\alpha \tilde{N}_v$, is shown as a function of the potential strength, $\alpha$.
As $\alpha$ is increased, waters are systematically displaced from the peptide hydration shell.
(c) The conformation free energy landscape of the peptide, $\beta F_\alpha(\Phi,\Psi)$, in the presence of the unfavorable potential is shown for three value of the potential strength, $\beta\alpha = 0, 2.2$, and $8$.
Snapshots of alanine dipeptide in select basins are also shown.
(d) The free energy barrier, $F_{\rm max}$, obtained by integrating $\exp[-\beta F_\alpha(\Phi,\Psi)]$
over $\Phi$ for $-4 \le \Phi \le 10$~degrees and all $\Psi$ is shown as a function of $\alpha$. 
$F_{\rm max}$ corresponds to the barrier for transitioning between right- and left-handed turns,
and is the lowest at $\beta\alpha\approx2$ when the peptide hydration shell is roughly half empty.
}
\label{fig:alanine}
\end{figure*}

\section{How the Hydration of a Flexible Solute Influences its Conformation}
%
The conformational landscape of a flexible solute arises from an interplay between its intra-molecular interactions,
 and its interactions with the solvent~\cite{sheng1994monte,Jayaraman}.
Consequently, as the hydration of a flexible solute is varied, and the balance between intramolecular and solute-water interactions is modulated, a change in solute conformation can be triggered~\cite{garde07,garde08,rodriguez2015mechanism}. 
We now illustrate the use of dynamic INDUS to characterize how the hydration of a prototypical flexible solute influences its conformational landscape.

\subsection{Alanine Dipeptide}
%
Peptides represent a versatile template for diverse biological and materials applications~\cite{Rossky:1999:BioPhys.J,Shell:2010:MS,Shea:2015:Langmuir}. 
In a first step towards understanding how the conformational ensemble of a peptide is modulated by its hydration,
here we study alanine dipeptide using dynamic INDUS.
Alanine dipeptide has been extensively studied using molecular simulations~\cite{Ala2-Levy-2015,Ala2-Levy-2004}, and often serves as a test-bed for enhanced sampling techniques.
Although the conformational free energy landscape of alanine dipeptide can be described using only
its two backbone torsional angles, $\Phi$ and $\Psi$ (Figure~\ref{fig:alanine}a),
the landscape, $F(\Phi,\Psi)$, captures many of the relevant conformations and transitions that occur in much larger peptides and proteins.

To modulate the hydration of alanine dipeptide using dynamic INDUS, we employ an observation volume 
that encompasses the first hydration shell of the peptide;
$v$ is defined as the union of spherical sub-volumes of radius, $R_v = 0.6$~nm, 
that are centered on each of the 10 peptide heavy atoms.
We then apply an unfavorable potential, $\alpha \Ntv$.
As the strength of the potential, $\alpha$, is increased, 
water is systematically displaced from $v$, 
resulting in a decrease in the average number of waters in $v$ (Figure~\ref{fig:alanine}b).
For each $\alpha$, we then use umbrella sampling to estimate the peptide conformational free energy landscape, $F_\alpha(\Phi,\Psi)$.
In Figure~\ref{fig:alanine}c, we plot $F_\alpha(\Phi,\Psi)$ for three select values of $\beta\alpha = 0, 2.2$, and $8$,
which correspond respectively to the peptide being fully hydrated, having roughly half its hydration waters, and being almost completely dewetted.
For $\beta\alpha = 0$, $F_\alpha(\Phi,\Psi)$ represents the landscape of the hydrated alanine dipeptide, 
whereas for $\beta\alpha = 8$, $F_\alpha(\Phi,\Psi)$ corresponds closely to the landscape of the peptide in vacuum.
As such, these landscapes display a number of well-characterized basins, which have discussed in detail elsewhere~\cite{Ala2-Levy-2015,Ala2-Levy-2004}.

Although relatively small barriers ($\lesssim 5 \kbt$) must be overcome for transitions between most basins,
a high free energy region ($\gtrsim 10 \kbt$) in the vicinity of $\Phi=0$ must be crossed 
to transition between a basin with $\Phi < 0$ and one with $\Phi > 0$.
In Figure~\ref{fig:alanine}c, we also show simulation snapshots of alanine dipeptide in individual basins, 
which are called $\alpha_{\rm{R}}$ ($\Phi < 0$) and $\alpha_{\rm{L}}$ ($\Phi > 0$), 
and are populated by right- and left-handed helices, respectively.
These snapshots along with a snapshot from the barrier region ($\Phi \approx 0$)
collectively highlight how the alanine dipeptide molecule rotates with increasing $\Phi$, 
as it transitions from a right-handed to a left-handed turn.
We find that the barrier height for transitioning between the turns is higher for the nearly 
dewetted peptide ($\beta\alpha = 8$) than it is for the full hydrated peptide ($\beta\alpha = 0$).
Interestingly, at intermediate hydration levels, the barrier appears to be lower than at $\beta\alpha = 0$ or $\beta\alpha = 8$.
To further investigate how the barrier height varies with hydration, we obtain estimates for 
the barrier free energy, $F_{\rm max}$, as a function of $\alpha$ 
by integrating $\exp[-\beta F_\alpha(\Phi,\Psi)]$ over the entire range of $\Psi$-values, 
and over $\Phi$ for $-4 \le \Phi \le 10$~degrees; 
the range of integration is highlighted using rectangular boxes in Figure~\ref{fig:alanine}c.
In Figure~\ref{fig:alanine}d, $\beta F_{\rm max}$ thus obtained is shown as a function of $\beta\alpha$, 
and confirms that there is an optimal hydration level that corresponds to the lowest barrier
for transitioning between the right- and left-handed turns.
%
Our results thus highlight that the barrier is the lowest for a partially hydrated peptide,
and suggest that a transition between the two turns may proceed through coupled fluctuations 
in the hydration of the peptide and its conformation.
They also suggest that the peptide may be able to transition more readily in environments that can facilitate the partial dewetting of its hydration shell, e.g., hydrophobic surfaces or confinement~\cite{anand2010conformational,Vembanur:2013,Shea:2015:Langmuir}.
%

\begin{figure}[htb]
\centering
\includegraphics[width=0.48\textwidth]{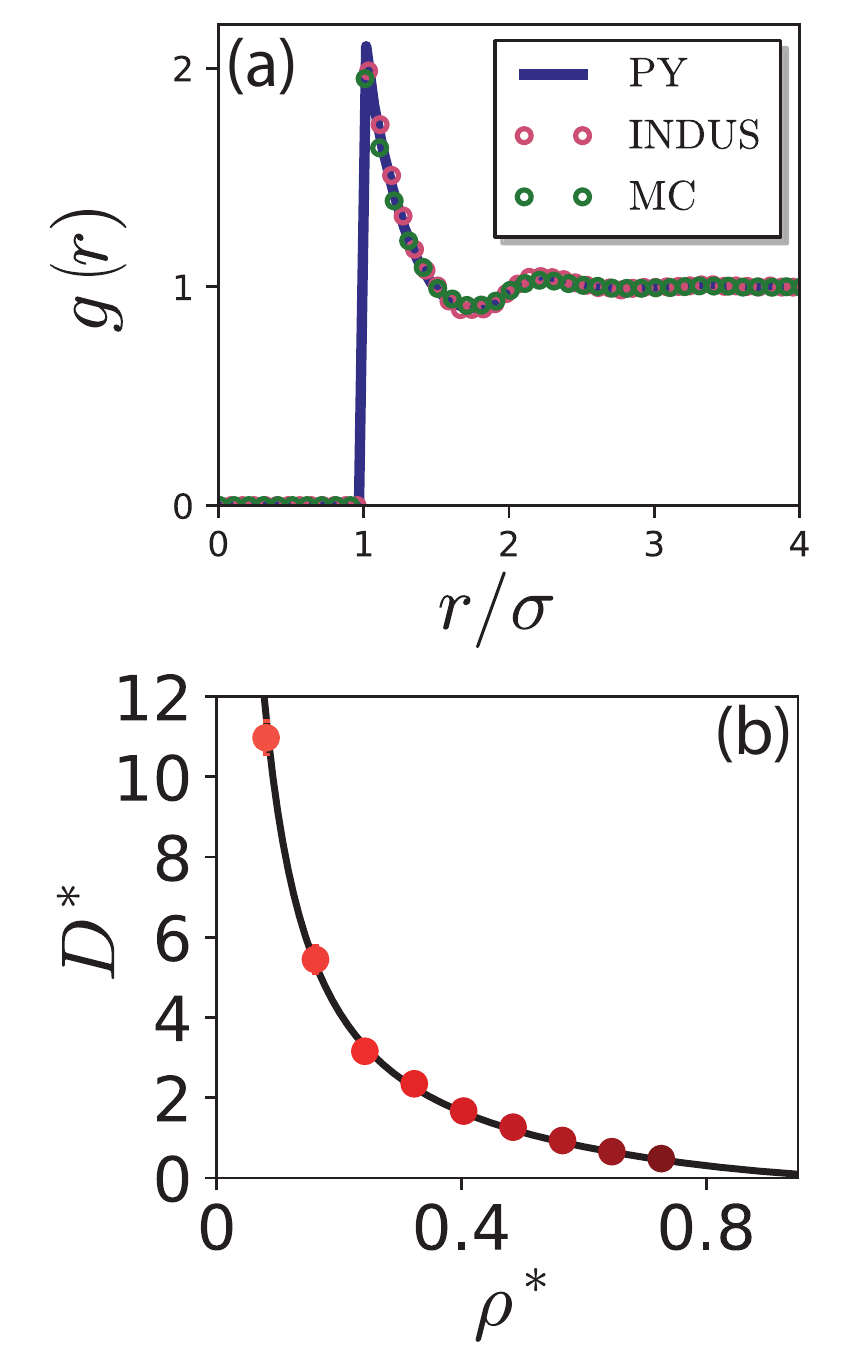} 
\caption{
(a) The radial distribution function, $g(r)$, for the mimic hard sphere system with a reduced density, {$\rho^* = 0.484$}, 
was obtained using dynamic INDUS, and is found to be in good agreement with the $g(r)$ of a true hard sphere fluid 
obtained from Monte Carlo simulations and the corrected Percus-Yevick equation~\cite{VerletWeis}.
(b) The dependence of the reduced diffusion coefficient, $D^*$, on the reduced density, $\rho^*$, 
obtained using dynamic INDUS simulations of mimic hard spheres (symbols) 
is in agreement with the empirical relation of Speedy (Equation~\ref{eq:speedy})~\cite{speedy}.
}
\label{fig:hs}
\end{figure}

\section{Mimic Hard Sphere Systems}
%
When the center of a spherical $v$ of radius $R_v$ is pegged to an ideal (non-interacting) particle,
and a sufficiently strong unfavorable biasing potential is applied to the water molecules in $v$,
waters are excluded from $v$, and configurations with $N_v = 0$ are obtained;
that is, the ideal particle pegged to the center of $v$ has nucleated a cavity (or a hard sphere) of radius, $R_v$.
If several such ideal particles are introduced in the system, each with its own (independent) unfavorable biasing potential,
which excludes not just waters, but also other ideal particles, a mixture of hard spheres (HS) and water can be obtained.
In the limit that the system only contains such ideal particles and no waters, it mimics a HS fluid.
To be precise, such a system corresponds to a soft analogue of a regularized HS fluid; 
configurations with $N_v > 0$ carry an energetic penalty that is high, but not infinite.
%

To mimic a HS system as described above, and to characterize its dynamic properties,
here we employ $M=1000$ ideal particles, which interact with one another only through the INDUS potential.
The system Hamiltonian is:
$$
\mathcal{H}(\{\rb_i\}) = \sum_{i=1}^{M} \frac{\kappa}{2} ( \Nt_i - N^*)^2,
$$
where $\Nt_i$ is the coarse-grained number of particles in a spherical observation volume, $v_i$,
which is centered on particle $i$, and has a radius of $R_v = 0.3$~nm.
The INDUS potential parameters were chosen to be $\kappa=1$~kJ/mol and $N^*=-10$.
The smoothed indicator functions were chosen to have a width of $\sigma=0.01$~nm and are cutoff at $r_{\rm c}=0.03$~nm.
As long as initial configurations were chosen without any overlap, 
these parameters were sufficient to exclude all particles $j\ne i$ from $v_i$ 
for all $i$ in all the configurations that were sampled.

%
We first characterize the structure of this mimic HS fluid by estimating its radial distribution function, $g(r)$, 
for a system with a reduced density $\rho^* \equiv \rho d^3 = 0.484$, where $\rho=M/V$ is the particle density, 
$V$ is the volume of the simulation box, and 
$d$ is the effective HS diameter.
To approximate $d$ for these particles, we consider the effect of the biasing potential in the limit of pairwise interactions, 
and use the corresponding pair potential, $u_0(r)$, between two particles as~\cite{blip,wca,Remsing:2013}
\begin{equation}
d=\int_0^\infty dr \brac{1-e^{-\beta u_0(r)}}.
\label{eq:d}
\end{equation}
This procedure yields $d=0.3166$~nm, which is slightly greater than $R_v = 0.3$~nm.
The radial distribution function, shown in Figure~\ref{fig:hs}a, displays good agreement with the $g(r)$ of a true HS fluid 
obtained from a Monte Carlo simulation, as well as the $g(r)$ predicted by the corrected Percus-Yevick equation of Verlet and Weis~\cite{VerletWeis}.

Because our mimic HS system is studied using MD simulations, 
we can readily estimate the diffusion coefficient of the mimic hard spheres
by fitting the long time behavior of the mean squared displacement to a straight line, 
following the Einstein relation:
$D=\lim_{t\rightarrow\infty} \avg{\len{\rb_i(t)-\rb_i(0)}^2} / 6t$,
where $\rb_i(t)$ is the position of particle $i$ at time $t$.
In this way, we estimate diffusion coefficients, $D$, for mimic HS systems with varying densities.
As shown in Figure~\ref{fig:hs}b, as particle density is increased, the diffusion coefficients decrease, 
as expected for HS systems.
We also compare our estimated dependence of $D$ on $\rho$ to the empirical function proposed by Speedy~\cite{speedy} for HS fluids,
\begin{equation}
D^*\equiv \frac{D}{D_0}
= \frac{1}{\rho^*} \bigg(1 - \frac{\rho^*}{1.09} \bigg) (1+ 0.4 \rho^{*2} - 0.83\rho^{*4}),
\label{eq:speedy}
\end{equation}
where
\begin{equation}
D_0 = \frac{3}{8}\para{\frac{\kbt d^2}{\pi m}}^{1/2}
\end{equation}
pertains to the zero density limit of the diffusion coefficient and $m$ is the mass of a HS particle.
As shown in Figure~\ref{fig:hs}b, the behavior of $D^*(\rho^*)$ for our mimic HS system 
is in excellent agreement with Equation~\ref{eq:speedy}, suggesting that
the dynamics of a true HS fluid are well-described by our mimic HS system.
Such a correspondence between soft and hard sphere systems follows from the WCA picture of simple liquid structure~\cite{wca},
and has also been used to accurately predict transport properties of simple liquids~\cite{Chandler:AccChemRes:1974}.

Although we have used dynamic INDUS to mimic hard sphere systems here,
rhe above approach can also be used to mimic other systems with discrete potentials, such as a square well fluid.
This can be accomplished by combining unfavorable particle-excluding INDUS potentials at one length scale ($R_v$) 
with favorable interactions that act on another, larger length scale, $\lambda R_v~(\lambda>1)$;
i.e., in addition to having particles excluded from a hard-core region, their presence could be favored in a second observation volume that serves as the attractive region~\cite{Remsing:JCP:2015}.

\section{Conclusions and Outlook}
%
%
The statistics of solvent density fluctuations play a central role in collective solvent-mediated phenomena,
including the important class of processes that are driven by hydrophobic effects.
The Indirect Umbrella Sampling (INDUS) method, 
which was developed to characterize such solvent fluctuations 
in observation volumes of all shapes and sizes,
has led to numerous insights on hydrophobic hydration and interactions.
However, within the INDUS framework, 
the positions of the observation volumes of interest as well as their shapes must be specified a priori, 
and must remain static throughout the biased simulations. 
As a result, it is not possible to use INDUS for characterizing the role of solvent fluctuations in dynamic processes, 
such as self-assembly or conformational change.
To address this challenge, here we generalize the INDUS method to sample solvent density fluctuations 
in dynamical observation volumes whose positions and shapes can evolve.
We focus primarily on observation volumes that are spherical or unions of spherical sub-volumes;
however, the approach presented here is fairly general, 
and the underlying ideas can be extended to volumes of other shapes, such as cuboids or cylinders, 
as well as unions and intersections of such sub-volumes.

In addition to outlining the methodological framework for generalizing INDUS to dynamical volumes, 
we also highlight the utility of the dynamic INDUS method using a number of illustrative examples.
First, we characterize the free energetics of water density fluctuations 
in dynamical volumes in bulk water using dummy particles to define $v$.
We find that it is easier to dewet a flexible (dynamic) oligomeric volume 
than it is to dewet the corresponding rigid (static) volume.
Next, we define the dynamical $v$ using real particles, which enables us to probe the 
hydration shells of these particles as they move with respect to one another.
By defining $v$ to be the union of the hydration shells of several methane molecules,
we are able capture the role of water in the association of these methanes.
We study both purely repulsive methanes and those with dispersive attractions, 
and find that in both cases, there is a strong correlation between desolvation and assembly.
Although we use dynamic INDUS to study the assembly of small hydrophobes, 
the method itself is fairly general, and could be used to study diverse assemblies, 
ranging from micelle formation to protein interactions.
Moreover, because desolvation barriers hinder diverse self-assemblies driven 
by hydrophobic interactions~\cite{sharma2012evaporation,sharma2012free,Vembanur:2013,Remsing:PNAS:2015},
the dynamic INDUS method will also facilitate the enhanced sampling of such assemblies by 
allowing for the biasing of the relevant (slow) solvent co-ordinates.

%
An important manifestation of the hydrophobic effect is solvent-mediated conformation change,
such as seen in the the folding of proteins or the collapse of hydrophobic polymers in water~\cite{Miller:PNAS:2007}.
Here we illustrate explore the coupling between the hydration and the conformational preferences of alanine dipeptide
by defining $v$ to be the dynamically evolving hydration shell of the peptide.
Interestingly, we find that the transition between right- and left-handed turns has the lowest barrier
when the peptide is partially dehydrated.
Although the free energetic landscapes of many conformationally flexible molecules
have been extensively characterized using enhanced sampling methods,  
previous studies have averaged over the solvent degrees of freedom to 
obtain free energy as a function of conformational order parameters, such as $\Rg$ or dihedral angles~\cite{garde07,garde08,Ala2-Levy-2015,Ala2-Levy-2004}.
The resulting loss of information makes it challenging to anticipate how the solute conformational landscape 
will respond to perturbations, such as proximity to interfaces or the addition of co-solutes.
By providing a framework for characterizing the free energetics of solvent fluctuations in the solvation shells of conformationally flexible solutes, we hope that our work will lead to a better understanding of the interplay between solvation and conformation for diverse classes of flexible molecules, ranging from peptides and peptido-mimics to polymers and unstructured regions of proteins.
We hope that the dynamic INDUS framework will also stimulate the development of sophisticated coarse-grained models~\cite{Mundy:JCP:2015, Noid:2015:JCP, Shell:2016:JCP,saunders_coarse-graining_2013}, which retain not only the essential solute degrees of freedom, but also the slow solvent co-ordinate degrees of freedom.

%
Finally, we illustrate how the dynamic INDUS method can be used to mimic hard particle systems. 
We use the method to simulate the dynamics of nearly-hard spheres using MD simulations, 
and estimate the corresponding diffusion coefficients.
The ideas underlying this approach could also be extended to particles that interact via square-well potentials, 
thereby enabling inclusion of discrete attractive interactions~\cite{Remsing:JCP:2015}.
By using collections of dummy particles to define $v$, it may also be possible to simulate the 
dynamics of non-spherical pseudo-hard particles with other shapes, such as rods or tetrahedra.

\acknowledgements
A.J.P. gratefully acknowledges financial support from the National Science Foundation through the University of Pennsylvania Materials Research Science and Engineering Center (NSF UPENN MRSEC DMR 1720530) and through grants CBET 1652646 and CHE 1665339, as well as a fellowship from the Alfred P. Sloan Research Foundation (FG-2017-9406). 
Z.J. was supported by the Charles E. Kaufman Foundation (KA-2015-79204) and the National Science Foundation (CBET 1511437).
N.B.R. was supported by the National Science Foundation (CBET 1652646).
A.J.P. thanks Adam Willard for insightful discussions that motivated this work.

\providecommand{\latin}[1]{#1}
\makeatletter
\providecommand{\doi}
  {\begingroup\let\do\@makeother\dospecials
  \catcode`\{=1 \catcode`\}=2 \doi@aux}
\providecommand{\doi@aux}[1]{\endgroup\texttt{#1}}
\makeatother
\providecommand*\mcitethebibliography{\thebibliography}
\csname @ifundefined\endcsname{endmcitethebibliography}
  {\let\endmcitethebibliography\endthebibliography}{}


\begin{mcitethebibliography}{94}
\providecommand*\natexlab[1]{#1}
\providecommand*\mciteSetBstSublistMode[1]{}
\providecommand*\mciteSetBstMaxWidthForm[2]{}
\providecommand*\mciteBstWouldAddEndPuncttrue
  {\def\EndOfBibitem{\unskip.}}
\providecommand*\mciteBstWouldAddEndPunctfalse
  {\let\EndOfBibitem\relax}
\providecommand*\mciteSetBstMidEndSepPunct[3]{}
\providecommand*\mciteSetBstSublistLabelBeginEnd[3]{}
\providecommand*\EndOfBibitem{}
\mciteSetBstSublistMode{f}
\mciteSetBstMaxWidthForm{subitem}{(\alph{mcitesubitemcount})}
\mciteSetBstSublistLabelBeginEnd
  {\mcitemaxwidthsubitemform\space}
  {\relax}
  {\relax}

\bibitem[Southall \latin{et~al.}(2002)Southall, Dill, and Haymet]{dill_rev02}
Southall,~N.~T.; Dill,~K.~A.; Haymet,~A. D.~J. A View of the Hydrophobic
  Effect. \emph{J. Phys. Chem. B} \textbf{2002}, \emph{106}, 521--533\relax
\mciteBstWouldAddEndPuncttrue
\mciteSetBstMidEndSepPunct{\mcitedefaultmidpunct}
{\mcitedefaultendpunct}{\mcitedefaultseppunct}\relax
\EndOfBibitem
\bibitem[Chandler(2005)]{Chandler:Nature:2005}
Chandler,~D. Interfaces and the driving force of hydrophobic assembly.
  \emph{Nature} \textbf{2005}, \emph{437}, 640--647\relax
\mciteBstWouldAddEndPuncttrue
\mciteSetBstMidEndSepPunct{\mcitedefaultmidpunct}
{\mcitedefaultendpunct}{\mcitedefaultseppunct}\relax
\EndOfBibitem
\bibitem[Rasaiah \latin{et~al.}(2008)Rasaiah, Garde, and Hummer]{garde_rev}
Rasaiah,~J.~C.; Garde,~S.; Hummer,~G. Water in Nonpolar Confinement: From
  Nanotubes to Proteins and Beyond. \emph{Ann. Rev. Phys. Chem.} \textbf{2008},
  \emph{59}, 713--740\relax
\mciteBstWouldAddEndPuncttrue
\mciteSetBstMidEndSepPunct{\mcitedefaultmidpunct}
{\mcitedefaultendpunct}{\mcitedefaultseppunct}\relax
\EndOfBibitem
\bibitem[Berne \latin{et~al.}(2009)Berne, Weeks, and Zhou]{berne_rev09}
Berne,~B.~J.; Weeks,~J.~D.; Zhou,~R. Dewetting and Hydrophobic Interaction in
  Physical and Biological Systems. \emph{Ann. Rev. Phys. Chem.} \textbf{2009},
  \emph{60}, 85--103\relax
\mciteBstWouldAddEndPuncttrue
\mciteSetBstMidEndSepPunct{\mcitedefaultmidpunct}
{\mcitedefaultendpunct}{\mcitedefaultseppunct}\relax
\EndOfBibitem
\bibitem[Jamadagni \latin{et~al.}(2011)Jamadagni, Godawat, and
  Garde]{Jamadagni:ARCB:2011}
Jamadagni,~S.~N.; Godawat,~R.; Garde,~S. Hydrophobicity of Proteins and
  Interfaces: Insights from Density Fluctuations. \emph{Ann. Rev. Chem. Biomol.
  Engg.} \textbf{2011}, \emph{2}, 147--171\relax
\mciteBstWouldAddEndPuncttrue
\mciteSetBstMidEndSepPunct{\mcitedefaultmidpunct}
{\mcitedefaultendpunct}{\mcitedefaultseppunct}\relax
\EndOfBibitem
\bibitem[Giovambattista \latin{et~al.}(2012)Giovambattista, Rossky, and
  Debenedetti]{Pablo_Review}
Giovambattista,~N.; Rossky,~P.; Debenedetti,~P. Computational Studies of
  Pressure, Temperature, and Surface Effects on the Structure and
  Thermodynamics of Confined Water. \emph{Annu. Rev. Phys. Chem.}
  \textbf{2012}, \emph{63}, 179--200\relax
\mciteBstWouldAddEndPuncttrue
\mciteSetBstMidEndSepPunct{\mcitedefaultmidpunct}
{\mcitedefaultendpunct}{\mcitedefaultseppunct}\relax
\EndOfBibitem
\bibitem[Maibaum \latin{et~al.}(2004)Maibaum, Dinner, and
  Chandler]{MaibaumDinnerChandler2004}
Maibaum,~L.; Dinner,~A.~R.; Chandler,~D. Micelle formation and the hydrophobic
  effect. \emph{J. Phys. Chem. B} \textbf{2004}, \emph{108}, 6778--6781\relax
\mciteBstWouldAddEndPuncttrue
\mciteSetBstMidEndSepPunct{\mcitedefaultmidpunct}
{\mcitedefaultendpunct}{\mcitedefaultseppunct}\relax
\EndOfBibitem
\bibitem[Meng and Ashbaugh(2013)Meng, and Ashbaugh]{meng2013pressure}
Meng,~B.; Ashbaugh,~H.~S. Pressure reentrant assembly: Direct simulation of
  volumes of micellization. \emph{Langmuir} \textbf{2013}, \emph{29},
  14743--14747\relax
\mciteBstWouldAddEndPuncttrue
\mciteSetBstMidEndSepPunct{\mcitedefaultmidpunct}
{\mcitedefaultendpunct}{\mcitedefaultseppunct}\relax
\EndOfBibitem
\bibitem[Rabani \latin{et~al.}(2003)Rabani, Reichman, Geissler, and
  Brus]{Rabani:2003}
Rabani,~E.; Reichman,~D.~R.; Geissler,~P.~L.; Brus,~L.~E. Drying-mediated
  self-assembly of nanoparticles. \emph{Nature} \textbf{2003}, \emph{426},
  271--274\relax
\mciteBstWouldAddEndPuncttrue
\mciteSetBstMidEndSepPunct{\mcitedefaultmidpunct}
{\mcitedefaultendpunct}{\mcitedefaultseppunct}\relax
\EndOfBibitem
\bibitem[Morrone \latin{et~al.}(2012)Morrone, Li, and Berne]{Morrone:2012}
Morrone,~J.~A.; Li,~J.; Berne,~B.~J. Interplay between Hydrodynamics and the
  Free Energy Surface in the Assembly of Nanoscale Hydrophobes. \emph{J. Phys.
  Chem. B} \textbf{2012}, \emph{116}, 378--389\relax
\mciteBstWouldAddEndPuncttrue
\mciteSetBstMidEndSepPunct{\mcitedefaultmidpunct}
{\mcitedefaultendpunct}{\mcitedefaultseppunct}\relax
\EndOfBibitem
\bibitem[Levy and Onuchic(2006)Levy, and Onuchic]{Levy:2006aa}
Levy,~Y.; Onuchic,~J.~N. Water mediation in protein folding and molecular
  recognition. \emph{Annu Rev Biophys Biomol Struct} \textbf{2006}, \emph{35},
  389--415\relax
\mciteBstWouldAddEndPuncttrue
\mciteSetBstMidEndSepPunct{\mcitedefaultmidpunct}
{\mcitedefaultendpunct}{\mcitedefaultseppunct}\relax
\EndOfBibitem
\bibitem[Dill and MacCallum(2012)Dill, and MacCallum]{Dill:2012aa}
Dill,~K.~A.; MacCallum,~J.~L. The protein-folding problem, 50 years on.
  \emph{Science} \textbf{2012}, \emph{338}, 1042--6\relax
\mciteBstWouldAddEndPuncttrue
\mciteSetBstMidEndSepPunct{\mcitedefaultmidpunct}
{\mcitedefaultendpunct}{\mcitedefaultseppunct}\relax
\EndOfBibitem
\bibitem[Krone \latin{et~al.}(2008)Krone, Hua, Soto, Zhou, Berne, and
  Shea]{shea08}
Krone,~M.~G.; Hua,~L.; Soto,~P.; Zhou,~R.; Berne,~B.~J.; Shea,~J.-E. Role of
  Water in Mediating the Assembly of Alzheimer Amyloid-beta Abeta16-22
  Protofilaments. \emph{J. Am. Chem. Soc.} \textbf{2008}, \emph{130},
  11066--11072\relax
\mciteBstWouldAddEndPuncttrue
\mciteSetBstMidEndSepPunct{\mcitedefaultmidpunct}
{\mcitedefaultendpunct}{\mcitedefaultseppunct}\relax
\EndOfBibitem
\bibitem[Thirumalai \latin{et~al.}(2012)Thirumalai, Reddy, and
  Straub]{Thirumalai:2012}
Thirumalai,~D.; Reddy,~G.; Straub,~J.~E. Role of Water in Protein Aggregation
  and Amyloid Polymorphism. \emph{Acc. Chem Res.} \textbf{2012}, \emph{45},
  83--92\relax
\mciteBstWouldAddEndPuncttrue
\mciteSetBstMidEndSepPunct{\mcitedefaultmidpunct}
{\mcitedefaultendpunct}{\mcitedefaultseppunct}\relax
\EndOfBibitem
\bibitem[Li and Walker(2011)Li, and Walker]{Li:PNAS:2011}
Li,~I.~T.; Walker,~G.~C. Signature of hydrophobic hydration in a single
  polymer. \emph{Proc. Natl. Acad. Sci. U.S.A.} \textbf{2011}, \emph{108},
  16527--16532\relax
\mciteBstWouldAddEndPuncttrue
\mciteSetBstMidEndSepPunct{\mcitedefaultmidpunct}
{\mcitedefaultendpunct}{\mcitedefaultseppunct}\relax
\EndOfBibitem
\bibitem[Davis \latin{et~al.}(2012)Davis, Gierszal, Wang, and
  Ben-Amotz]{davis2012water}
Davis,~J.~G.; Gierszal,~K.~P.; Wang,~P.; Ben-Amotz,~D. Water structural
  transformation at molecular hydrophobic interfaces. \emph{Nature}
  \textbf{2012}, \emph{491}, 582\relax
\mciteBstWouldAddEndPuncttrue
\mciteSetBstMidEndSepPunct{\mcitedefaultmidpunct}
{\mcitedefaultendpunct}{\mcitedefaultseppunct}\relax
\EndOfBibitem
\bibitem[Ma \latin{et~al.}(2015)Ma, Wang, Acevedo-V{\'e}lez, Gellman, and
  Abbott]{ma2015modulation}
Ma,~C.~D.; Wang,~C.; Acevedo-V{\'e}lez,~C.; Gellman,~S.~H.; Abbott,~N.~L.
  Modulation of hydrophobic interactions by proximally immobilized ions.
  \emph{Nature} \textbf{2015}, \emph{517}, 347\relax
\mciteBstWouldAddEndPuncttrue
\mciteSetBstMidEndSepPunct{\mcitedefaultmidpunct}
{\mcitedefaultendpunct}{\mcitedefaultseppunct}\relax
\EndOfBibitem
\bibitem[Tang \latin{et~al.}(2017)Tang, Barnett, Gibb, and
  Ashbaugh]{tang2017guest}
Tang,~D.; Barnett,~J.~W.; Gibb,~B.~C.; Ashbaugh,~H.~S. Guest Controlled
  Nonmonotonic Deep Cavity Cavitand Assembly State Switching. \emph{J. Phys.
  Chem. B} \textbf{2017}, \emph{121}, 10717--10725\relax
\mciteBstWouldAddEndPuncttrue
\mciteSetBstMidEndSepPunct{\mcitedefaultmidpunct}
{\mcitedefaultendpunct}{\mcitedefaultseppunct}\relax
\EndOfBibitem
\bibitem[Remsing and Patel(2015)Remsing, and Patel]{Remsing:JCP:2015}
Remsing,~R.~C.; Patel,~A.~J. Water Density Fluctuations Relevant to Hydrophobic
  Hydration are Unaltered by Attractions. \emph{J. Chem. Phys.} \textbf{2015},
  \emph{142}, 024502\relax
\mciteBstWouldAddEndPuncttrue
\mciteSetBstMidEndSepPunct{\mcitedefaultmidpunct}
{\mcitedefaultendpunct}{\mcitedefaultseppunct}\relax
\EndOfBibitem
\bibitem[Hummer \latin{et~al.}(1996)Hummer, Garde, Garcia, Pohorille, and
  Pratt]{Hummer:PNAS:1996}
Hummer,~G.; Garde,~S.; Garcia,~A.~E.; Pohorille,~A.; Pratt,~L.~R. An
  information theory model of hydrophobic interactions. \emph{Proc. Nat. Acad.
  Sci.} \textbf{1996}, \emph{93}, 8951--8955\relax
\mciteBstWouldAddEndPuncttrue
\mciteSetBstMidEndSepPunct{\mcitedefaultmidpunct}
{\mcitedefaultendpunct}{\mcitedefaultseppunct}\relax
\EndOfBibitem
\bibitem[Huang \latin{et~al.}(2001)Huang, Geissler, and Chandler]{HGC}
Huang,~D.~M.; Geissler,~P.~L.; Chandler,~D. Scaling of hydrophobic free
  energies. \emph{J. Phys. Chem. B} \textbf{2001}, \emph{105}, 6704--6709\relax
\mciteBstWouldAddEndPuncttrue
\mciteSetBstMidEndSepPunct{\mcitedefaultmidpunct}
{\mcitedefaultendpunct}{\mcitedefaultseppunct}\relax
\EndOfBibitem
\bibitem[Patel \latin{et~al.}(2010)Patel, Varilly, and
  Chandler]{Patel:JPCB:2010}
Patel,~A.~J.; Varilly,~P.; Chandler,~D. Fluctuations of Water near Extended
  Hydrophobic and Hydrophilic Surfaces. \emph{J. Phys. Chem. B} \textbf{2010},
  \emph{114}, 1632 -- 1637\relax
\mciteBstWouldAddEndPuncttrue
\mciteSetBstMidEndSepPunct{\mcitedefaultmidpunct}
{\mcitedefaultendpunct}{\mcitedefaultseppunct}\relax
\EndOfBibitem
\bibitem[Remsing and Weeks(2013)Remsing, and Weeks]{Remsing:2013}
Remsing,~R.~C.; Weeks,~J.~D. Dissecting Hydrophobic Hydration and Association.
  \emph{J. Phys. Chem. B} \textbf{2013}, \emph{117}, 15479--15491\relax
\mciteBstWouldAddEndPuncttrue
\mciteSetBstMidEndSepPunct{\mcitedefaultmidpunct}
{\mcitedefaultendpunct}{\mcitedefaultseppunct}\relax
\EndOfBibitem
\bibitem[Patel and Garde(2014)Patel, and Garde]{Patel:JPCB:2014}
Patel,~A.~J.; Garde,~S. Efficient Method To Characterize the Context-Dependent
  Hydrophobicity of Proteins. \emph{J. Phys. Chem. B} \textbf{2014},
  \emph{118}, 1564--1573\relax
\mciteBstWouldAddEndPuncttrue
\mciteSetBstMidEndSepPunct{\mcitedefaultmidpunct}
{\mcitedefaultendpunct}{\mcitedefaultseppunct}\relax
\EndOfBibitem
\bibitem[Stillinger(1973)]{FHS:1973}
Stillinger,~F.~H. Structure in aqueous solutions of nonpolar solutes from the
  standpoint of scaled-particle theory. \emph{J. Solution Chem.} \textbf{1973},
  \emph{2}, 141--158\relax
\mciteBstWouldAddEndPuncttrue
\mciteSetBstMidEndSepPunct{\mcitedefaultmidpunct}
{\mcitedefaultendpunct}{\mcitedefaultseppunct}\relax
\EndOfBibitem
\bibitem[Pratt and Chandler(1977)Pratt, and Chandler]{Pratt:JCP:1977}
Pratt,~L.~R.; Chandler,~D. Theory of the hydrophobic effect. \emph{J. Chem.
  Phys.} \textbf{1977}, \emph{67}, 3683--3704\relax
\mciteBstWouldAddEndPuncttrue
\mciteSetBstMidEndSepPunct{\mcitedefaultmidpunct}
{\mcitedefaultendpunct}{\mcitedefaultseppunct}\relax
\EndOfBibitem
\bibitem[Ashbaugh and Pratt(2006)Ashbaugh, and Pratt]{ashbaugh_SPT}
Ashbaugh,~H.~S.; Pratt,~L.~R. Colloquium: Scaled particle theory and the length
  scales of hydrophobicity. \emph{Rev. Mod. Phys.} \textbf{2006}, \emph{78},
  159\relax
\mciteBstWouldAddEndPuncttrue
\mciteSetBstMidEndSepPunct{\mcitedefaultmidpunct}
{\mcitedefaultendpunct}{\mcitedefaultseppunct}\relax
\EndOfBibitem
\bibitem[Lum \latin{et~al.}(1999)Lum, Chandler, and Weeks]{LCW}
Lum,~K.; Chandler,~D.; Weeks,~J.~D. Hydrophobicity at Small and Large Length
  Scales. \emph{J. Phys. Chem. B} \textbf{1999}, \emph{103}, 4570--4577\relax
\mciteBstWouldAddEndPuncttrue
\mciteSetBstMidEndSepPunct{\mcitedefaultmidpunct}
{\mcitedefaultendpunct}{\mcitedefaultseppunct}\relax
\EndOfBibitem
\bibitem[Widom(1963)]{Widom:JCP:1963}
Widom,~B. Some topics in the theory of fluids. \emph{J. Chem. Phys.}
  \textbf{1963}, \emph{39}, 2808 -- 2812\relax
\mciteBstWouldAddEndPuncttrue
\mciteSetBstMidEndSepPunct{\mcitedefaultmidpunct}
{\mcitedefaultendpunct}{\mcitedefaultseppunct}\relax
\EndOfBibitem
\bibitem[Garde \latin{et~al.}(1996)Garde, Hummer, Garcia, Paulaitis, and
  Pratt]{Garde:PRL:1996}
Garde,~S.; Hummer,~G.; Garcia,~A.~E.; Paulaitis,~M.~E.; Pratt,~L.~R. Origin of
  Entropy Convergence in Hydrophobic Hydration and Protein Folding. \emph{Phys.
  Rev. Lett.} \textbf{1996}, \emph{77}, 4966--4968\relax
\mciteBstWouldAddEndPuncttrue
\mciteSetBstMidEndSepPunct{\mcitedefaultmidpunct}
{\mcitedefaultendpunct}{\mcitedefaultseppunct}\relax
\EndOfBibitem
\bibitem[Hummer \latin{et~al.}(1998)Hummer, Garde, Garcia, Paulaitis, and
  Pratt]{Hummer:PNAS:1998}
Hummer,~G.; Garde,~S.; Garcia,~A.; Paulaitis,~M.; Pratt,~L. The pressure
  dependence of hydrophobic interactions is consistent with the observed
  pressure denaturation of proteins. \emph{Proc. Natl. Acad. Sci. USA}
  \textbf{1998}, \emph{95}, 1552--1555\relax
\mciteBstWouldAddEndPuncttrue
\mciteSetBstMidEndSepPunct{\mcitedefaultmidpunct}
{\mcitedefaultendpunct}{\mcitedefaultseppunct}\relax
\EndOfBibitem
\bibitem[Varilly \latin{et~al.}(2011)Varilly, Patel, and Chandler]{LLCW}
Varilly,~P.; Patel,~A.~J.; Chandler,~D. An improved coarse-grained model of
  solvation and the hydrophobic effect. \emph{J. Chem. Phys.} \textbf{2011},
  \emph{134}, 074109\relax
\mciteBstWouldAddEndPuncttrue
\mciteSetBstMidEndSepPunct{\mcitedefaultmidpunct}
{\mcitedefaultendpunct}{\mcitedefaultseppunct}\relax
\EndOfBibitem
\bibitem[Vaikuntanathan and Geissler(2014)Vaikuntanathan, and
  Geissler]{Geissler:2014:PRL}
Vaikuntanathan,~S.; Geissler,~P.~L. Putting Water on a Lattice: The Importance
  of Long Wavelength Density Fluctuations in Theories of Hydrophobic and
  Interfacial Phenomena. \emph{Phys. Rev. Lett.} \textbf{2014}, \emph{112},
  020603\relax
\mciteBstWouldAddEndPuncttrue
\mciteSetBstMidEndSepPunct{\mcitedefaultmidpunct}
{\mcitedefaultendpunct}{\mcitedefaultseppunct}\relax
\EndOfBibitem
\bibitem[Vaikuntanathan \latin{et~al.}(2016)Vaikuntanathan, Rotskoff, Hudson,
  and Geissler]{Geissler:2016:PNAS}
Vaikuntanathan,~S.; Rotskoff,~G.; Hudson,~A.; Geissler,~P.~L. Necessity of
  capillary modes in a minimal model of nanoscale hydrophobic solvation.
  \emph{Proceedings of the National Academy of Sciences} \textbf{2016},
  \emph{113}, E2224--E2230\relax
\mciteBstWouldAddEndPuncttrue
\mciteSetBstMidEndSepPunct{\mcitedefaultmidpunct}
{\mcitedefaultendpunct}{\mcitedefaultseppunct}\relax
\EndOfBibitem
\bibitem[Xi and Patel(2016)Xi, and Patel]{Xi:PNAS:2016}
Xi,~E.; Patel,~A.~J. The Hydrophobic Effect and Fluctuations: The Long and the
  Short of it. \emph{Proc. Natl. Acad. Sci. U.S.A.} \textbf{2016},
  \emph{113}\relax
\mciteBstWouldAddEndPuncttrue
\mciteSetBstMidEndSepPunct{\mcitedefaultmidpunct}
{\mcitedefaultendpunct}{\mcitedefaultseppunct}\relax
\EndOfBibitem
\bibitem[Patel \latin{et~al.}(2011)Patel, Varilly, Chandler, and
  Garde]{Patel:JSP:2011}
Patel,~A.~J.; Varilly,~P.; Chandler,~D.; Garde,~S. Quantifying density
  fluctuations in volumes of all shapes and sizes using indirect umbrella
  sampling. \emph{J. Stat. Phys.} \textbf{2011}, \emph{145}, 265 -- 275\relax
\mciteBstWouldAddEndPuncttrue
\mciteSetBstMidEndSepPunct{\mcitedefaultmidpunct}
{\mcitedefaultendpunct}{\mcitedefaultseppunct}\relax
\EndOfBibitem
\bibitem[Patel \latin{et~al.}(2012)Patel, Varilly, Jamadagni, Hagan, Chandler,
  and Garde]{Patel:JPCB:2012}
Patel,~A.~J.; Varilly,~P.; Jamadagni,~S.~N.; Hagan,~M.~F.; Chandler,~D.;
  Garde,~S. Sitting at the Edge: How Biomolecules Use Hydrophobicity to Tune
  their Interactions and Function. \emph{J. Phys. Chem. B} \textbf{2012},
  \emph{116}, 2498 -- 2503\relax
\mciteBstWouldAddEndPuncttrue
\mciteSetBstMidEndSepPunct{\mcitedefaultmidpunct}
{\mcitedefaultendpunct}{\mcitedefaultseppunct}\relax
\EndOfBibitem
\bibitem[Zhou \latin{et~al.}(2004)Zhou, Huang, Margulis, and Berne]{berne04}
Zhou,~R.; Huang,~X.; Margulis,~C.~J.; Berne,~B.~J. Hydrophobic Collapse in
  Multidomain Protein Folding. \emph{Science} \textbf{2004}, \emph{305},
  1605--1609\relax
\mciteBstWouldAddEndPuncttrue
\mciteSetBstMidEndSepPunct{\mcitedefaultmidpunct}
{\mcitedefaultendpunct}{\mcitedefaultseppunct}\relax
\EndOfBibitem
\bibitem[Liu \latin{et~al.}(2005)Liu, Huang, Zhou, and Berne]{berne05_melittin}
Liu,~P.; Huang,~X.; Zhou,~R.; Berne,~B.~J. Observation of a dewetting
  transition in the collapse of the melittin tetramer. \emph{Nature}
  \textbf{2005}, \emph{437}, 159--162\relax
\mciteBstWouldAddEndPuncttrue
\mciteSetBstMidEndSepPunct{\mcitedefaultmidpunct}
{\mcitedefaultendpunct}{\mcitedefaultseppunct}\relax
\EndOfBibitem
\bibitem[Choudhury and Pettitt(2007)Choudhury, and Pettitt]{chou_dewet}
Choudhury,~N.; Pettitt,~B.~M. The Dewetting Transition and The Hydrophobic
  Effect. \emph{J. Am. Chem. Soc.} \textbf{2007}, \emph{129}, 4847--4852\relax
\mciteBstWouldAddEndPuncttrue
\mciteSetBstMidEndSepPunct{\mcitedefaultmidpunct}
{\mcitedefaultendpunct}{\mcitedefaultseppunct}\relax
\EndOfBibitem
\bibitem[Patel \latin{et~al.}(2011)Patel, Varilly, Jamadagni, Acharya, Garde,
  and Chandler]{Patel:PNAS:2011}
Patel,~A.~J.; Varilly,~P.; Jamadagni,~S.~N.; Acharya,~H.; Garde,~S.;
  Chandler,~D. Extended surfaces modulate hydrophobic interactions of
  neighboring solutes. \emph{Proc. Natl. Acad. Sci. U.S.A.} \textbf{2011},
  \emph{108}, 17678 -- 17683\relax
\mciteBstWouldAddEndPuncttrue
\mciteSetBstMidEndSepPunct{\mcitedefaultmidpunct}
{\mcitedefaultendpunct}{\mcitedefaultseppunct}\relax
\EndOfBibitem
\bibitem[Vembanur \latin{et~al.}(2013)Vembanur, Patel, Sarupria, and
  Garde]{Vembanur:2013}
Vembanur,~S.; Patel,~A.~J.; Sarupria,~S.; Garde,~S. On the Thermodynamics and
  Kinetics of Hydrophobic Interactions at Interfaces. \emph{J. Phys. Chem. B}
  \textbf{2013}, \emph{117}, 10261--10270\relax
\mciteBstWouldAddEndPuncttrue
\mciteSetBstMidEndSepPunct{\mcitedefaultmidpunct}
{\mcitedefaultendpunct}{\mcitedefaultseppunct}\relax
\EndOfBibitem
\bibitem[Pohorille \latin{et~al.}(2010)Pohorille, Jarzynski, and
  Chipot]{Jarzynski:2010:JPCB}
Pohorille,~A.; Jarzynski,~C.; Chipot,~C. Good practices in free-energy
  calculations. \emph{J. Phys. Chem. B} \textbf{2010}, \emph{114},
  10235--10253\relax
\mciteBstWouldAddEndPuncttrue
\mciteSetBstMidEndSepPunct{\mcitedefaultmidpunct}
{\mcitedefaultendpunct}{\mcitedefaultseppunct}\relax
\EndOfBibitem
\bibitem[Xi \latin{et~al.}(2016)Xi, Remsing, and Patel]{Xi:JCTC:2016}
Xi,~E.; Remsing,~R.~C.; Patel,~A.~J. Sparse Sampling of Water Density
  Fluctuations in Interfacial Environments. \emph{J. Chem. Theory Comput.}
  \textbf{2016}, \emph{12}, 706--713\relax
\mciteBstWouldAddEndPuncttrue
\mciteSetBstMidEndSepPunct{\mcitedefaultmidpunct}
{\mcitedefaultendpunct}{\mcitedefaultseppunct}\relax
\EndOfBibitem
\bibitem[Xi \latin{et~al.}(2018)Xi, Marks, Fialoke, and Patel]{Xi:Molsim:2018}
Xi,~E.; Marks,~S.~M.; Fialoke,~S.; Patel,~A.~J. Sparse Sampling of Water
  Density Fluctuations Near Liquid-Vapor Coexistence. \emph{Mol. Simulation}
  \textbf{2018}, \emph{44}, 1124--1135\relax
\mciteBstWouldAddEndPuncttrue
\mciteSetBstMidEndSepPunct{\mcitedefaultmidpunct}
{\mcitedefaultendpunct}{\mcitedefaultseppunct}\relax
\EndOfBibitem
\bibitem[Wu and Garde(2014)Wu, and Garde]{wu2014lengthscale}
Wu,~E.; Garde,~S. Lengthscale-Dependent Solvation and Density Fluctuations in
  n-Octane. \emph{J. Phys. Chem. B} \textbf{2014}, \emph{119}, 9287--9294\relax
\mciteBstWouldAddEndPuncttrue
\mciteSetBstMidEndSepPunct{\mcitedefaultmidpunct}
{\mcitedefaultendpunct}{\mcitedefaultseppunct}\relax
\EndOfBibitem
\bibitem[Limmer \latin{et~al.}(2013)Limmer, Willard, Madden, and
  Chandler]{limmer2013hydration}
Limmer,~D.~T.; Willard,~A.~P.; Madden,~P.; Chandler,~D. Hydration of metal
  surfaces can be dynamically heterogeneous and hydrophobic. \emph{Proc. Natl.
  Acad. Sci. U.S.A.} \textbf{2013}, \emph{110}, 4200--4205\relax
\mciteBstWouldAddEndPuncttrue
\mciteSetBstMidEndSepPunct{\mcitedefaultmidpunct}
{\mcitedefaultendpunct}{\mcitedefaultseppunct}\relax
\EndOfBibitem
\bibitem[Rotenberg \latin{et~al.}(2011)Rotenberg, Patel, and
  Chandler]{Rotenberg:JACS:2011}
Rotenberg,~B.; Patel,~A.~J.; Chandler,~D. Molecular Explanation for Why Talc
  Surfaces can be Both Hydrophilic and Hydrophobic. \emph{J. Am. Chem. Soc.}
  \textbf{2011}, \emph{133}, 20521 -- 20527\relax
\mciteBstWouldAddEndPuncttrue
\mciteSetBstMidEndSepPunct{\mcitedefaultmidpunct}
{\mcitedefaultendpunct}{\mcitedefaultseppunct}\relax
\EndOfBibitem
\bibitem[Remsing \latin{et~al.}(2018)Remsing, Xi, and Patel]{Remsing:JPCB:2018}
Remsing,~R.~C.; Xi,~E.; Patel,~A.~J. Protein {{Hydration Thermodynamics}}:
  {{The Influence}} of {{Flexibility}} and {{Salt}} on {{Hydrophobin II
  Hydration}}. \emph{J. Phys. Chem. B} \textbf{2018}, \emph{122},
  3635--3646\relax
\mciteBstWouldAddEndPuncttrue
\mciteSetBstMidEndSepPunct{\mcitedefaultmidpunct}
{\mcitedefaultendpunct}{\mcitedefaultseppunct}\relax
\EndOfBibitem
\bibitem[Yu and Hagan(2012)Yu, and Hagan]{yu2012simulations}
Yu,~N.; Hagan,~M.~F. Simulations of HIV capsid protein dimerization reveal the
  effect of chemistry and topography on the mechanism of hydrophobic protein
  association. \emph{Biophys. J.} \textbf{2012}, \emph{103}, 1363--1369\relax
\mciteBstWouldAddEndPuncttrue
\mciteSetBstMidEndSepPunct{\mcitedefaultmidpunct}
{\mcitedefaultendpunct}{\mcitedefaultseppunct}\relax
\EndOfBibitem
\bibitem[Remsing \latin{et~al.}(2015)Remsing, Xi, Ranganathan, Sharma,
  Debenedetti, Garde, and Patel]{Remsing:PNAS:2015}
Remsing,~R.~C.; Xi,~E.; Ranganathan,~S.; Sharma,~S.; Debenedetti,~P.~G.;
  Garde,~S.; Patel,~A.~J. Pathways to dewetting in hydrophobic confinement.
  \emph{Proc. Natl. Acad. Sci. U.S.A.} \textbf{2015}, \emph{112},
  8181--8186\relax
\mciteBstWouldAddEndPuncttrue
\mciteSetBstMidEndSepPunct{\mcitedefaultmidpunct}
{\mcitedefaultendpunct}{\mcitedefaultseppunct}\relax
\EndOfBibitem
\bibitem[Prakash \latin{et~al.}(2016)Prakash, Xi, and Patel]{Prakash:PNAS:2016}
Prakash,~S.; Xi,~E.; Patel,~A.~J. Spontaneous Recovery of Superhydrophobicity
  on Nanotextured Surfaces. \emph{Proc. Natl. Acad. Sci. U.S.A.} \textbf{2016},
  \emph{113}, 5508--5513\relax
\mciteBstWouldAddEndPuncttrue
\mciteSetBstMidEndSepPunct{\mcitedefaultmidpunct}
{\mcitedefaultendpunct}{\mcitedefaultseppunct}\relax
\EndOfBibitem
\bibitem[Hess \latin{et~al.}(2008)Hess, Kutzner, van~der Spoel, and
  Lindahl]{gmx4ref}
Hess,~B.; Kutzner,~C.; van~der Spoel,~D.; Lindahl,~E. GROMACS 4: Algorithms for
  Highly Efficient, Load-Balanced, and Scalable Molecular Simulation. \emph{J.
  Chem. Theory Comp.} \textbf{2008}, 435 -- 447\relax
\mciteBstWouldAddEndPuncttrue
\mciteSetBstMidEndSepPunct{\mcitedefaultmidpunct}
{\mcitedefaultendpunct}{\mcitedefaultseppunct}\relax
\EndOfBibitem
\bibitem[Frenkel and Smit(2002)Frenkel, and Smit]{Frenkel_Smit}
Frenkel,~D.; Smit,~B. \emph{Understanding Molecular Simulations: From
  Algorithms to Applications}, 2nd ed.; Academic Press, New York, 2002\relax
\mciteBstWouldAddEndPuncttrue
\mciteSetBstMidEndSepPunct{\mcitedefaultmidpunct}
{\mcitedefaultendpunct}{\mcitedefaultseppunct}\relax
\EndOfBibitem
\bibitem[Berendsen \latin{et~al.}(1987)Berendsen, Grigera, and Straatsma]{SPCE}
Berendsen,~H. J.~C.; Grigera,~J.~R.; Straatsma,~T.~P. The Missing Term in
  Effective Pair Potentials. \emph{J. Phys. Chem.} \textbf{1987}, \emph{91},
  6269--6271\relax
\mciteBstWouldAddEndPuncttrue
\mciteSetBstMidEndSepPunct{\mcitedefaultmidpunct}
{\mcitedefaultendpunct}{\mcitedefaultseppunct}\relax
\EndOfBibitem
\bibitem[Ryckaert \latin{et~al.}(1977)Ryckaert, Ciccotti, and Berendsen]{SHAKE}
Ryckaert,~J.-P.; Ciccotti,~G.; Berendsen,~H. J.~C. Numerical integration of the
  cartesian equations of motion of a system with constraints: molecular
  dynamics of n-alkanes. \emph{J. Comp. Phys.} \textbf{1977}, \emph{23}, 327 --
  341\relax
\mciteBstWouldAddEndPuncttrue
\mciteSetBstMidEndSepPunct{\mcitedefaultmidpunct}
{\mcitedefaultendpunct}{\mcitedefaultseppunct}\relax
\EndOfBibitem
\bibitem[Essmann \latin{et~al.}(1995)Essmann, Perera, Berkowitz, Darden, Lee,
  and Pedersen]{PME}
Essmann,~U.; Perera,~L.; Berkowitz,~M.~L.; Darden,~T.; Lee,~H.; Pedersen,~L.~G.
  A smooth particle mesh Ewald method. \emph{J. Chem. Phys.} \textbf{1995},
  \emph{103}, 8577--8593\relax
\mciteBstWouldAddEndPuncttrue
\mciteSetBstMidEndSepPunct{\mcitedefaultmidpunct}
{\mcitedefaultendpunct}{\mcitedefaultseppunct}\relax
\EndOfBibitem
\bibitem[Bussi \latin{et~al.}(2007)Bussi, Donadio, and
  Parrinello]{Bussi:JCP:2007}
Bussi,~G.; Donadio,~D.; Parrinello,~M. Canonical sampling through velocity
  rescaling. \emph{J. Chem. Phys.} \textbf{2007}, \emph{126}, 014101\relax
\mciteBstWouldAddEndPuncttrue
\mciteSetBstMidEndSepPunct{\mcitedefaultmidpunct}
{\mcitedefaultendpunct}{\mcitedefaultseppunct}\relax
\EndOfBibitem
\bibitem[Parrinello and Rahman(1981)Parrinello, and Rahman]{Parrinello-Rahman}
Parrinello,~M.; Rahman,~A. Polymorphic transitions in single crystals: A new
  molecular dynamics method. \emph{J. Applied Phys.} \textbf{1981}, \emph{52},
  7182--7190\relax
\mciteBstWouldAddEndPuncttrue
\mciteSetBstMidEndSepPunct{\mcitedefaultmidpunct}
{\mcitedefaultendpunct}{\mcitedefaultseppunct}\relax
\EndOfBibitem
\bibitem[Martin and Siepmann(1998)Martin, and Siepmann]{trappe-ua}
Martin,~M.~G.; Siepmann,~J.~I. Transferable potentials for phase equilibria. 1.
  United-atom description of n-alkanes. \emph{J. Phys. Chem. B} \textbf{1998},
  \emph{102}, 2569--2577\relax
\mciteBstWouldAddEndPuncttrue
\mciteSetBstMidEndSepPunct{\mcitedefaultmidpunct}
{\mcitedefaultendpunct}{\mcitedefaultseppunct}\relax
\EndOfBibitem
\bibitem[Weeks \latin{et~al.}(1971)Weeks, Chandler, and Andersen]{wca}
Weeks,~J.~D.; Chandler,~D.; Andersen,~H.~C. Role of repulsive forces in forming
  the equilibrium structure of simple liquids. \emph{J. Chem. Phys.}
  \textbf{1971}, \emph{54}, 5237--5247\relax
\mciteBstWouldAddEndPuncttrue
\mciteSetBstMidEndSepPunct{\mcitedefaultmidpunct}
{\mcitedefaultendpunct}{\mcitedefaultseppunct}\relax
\EndOfBibitem
\bibitem[Cornell \latin{et~al.}(1995)Cornell, Cieplak, Bayly, Gould, Merz,
  Ferguson, Spellmeyer, Fox, Caldwell, and Kollman]{AMBER}
Cornell,~W.~D.; Cieplak,~P.; Bayly,~C.~I.; Gould,~I.~R.; Merz,~K.~M.;
  Ferguson,~D.~M.; Spellmeyer,~D.~C.; Fox,~T.; Caldwell,~J.~W.; Kollman,~P.~A.
  A Second Generation Force Field for the Simulation of Proteins, Nucleic
  Acids, and Organic Molecules. \emph{J. Am. Chem. Soc.} \textbf{1995},
  \emph{117}, 5179--5197\relax
\mciteBstWouldAddEndPuncttrue
\mciteSetBstMidEndSepPunct{\mcitedefaultmidpunct}
{\mcitedefaultendpunct}{\mcitedefaultseppunct}\relax
\EndOfBibitem
\bibitem[Lange \latin{et~al.}(2010)Lange, van~der Spoel, and
  de~Groot]{Lange_amber99sb}
Lange,~O.~F.; van~der Spoel,~D.; de~Groot,~B.~L. Scrutinizing Molecular
  Mechanics Force Fields on the Submicrosecond Timescale with NMR Data.
  \emph{Biophys. J.} \textbf{2010}, \emph{99}, 647 -- 655\relax
\mciteBstWouldAddEndPuncttrue
\mciteSetBstMidEndSepPunct{\mcitedefaultmidpunct}
{\mcitedefaultendpunct}{\mcitedefaultseppunct}\relax
\EndOfBibitem
\bibitem[Tan \latin{et~al.}(2012)Tan, Gallichio, Lapelosa, and Levy]{UWHAM}
Tan,~Z.; Gallichio,~E.; Lapelosa,~M.; Levy,~R.~M. Theory of binless multi-state
  free energy estimation with applications to protein-ligand binding. \emph{J.
  Chem. Phys.} \textbf{2012}, \emph{136}, 144102\relax
\mciteBstWouldAddEndPuncttrue
\mciteSetBstMidEndSepPunct{\mcitedefaultmidpunct}
{\mcitedefaultendpunct}{\mcitedefaultseppunct}\relax
\EndOfBibitem
\bibitem[Shirts and Chodera(2008)Shirts, and Chodera]{MBAR}
Shirts,~M.~R.; Chodera,~J.~D. Statistically optimal analysis of samples from
  multiple equilibrium states. \emph{J. Chem. Phys.} \textbf{2008}, \emph{129},
  124105\relax
\mciteBstWouldAddEndPuncttrue
\mciteSetBstMidEndSepPunct{\mcitedefaultmidpunct}
{\mcitedefaultendpunct}{\mcitedefaultseppunct}\relax
\EndOfBibitem
\bibitem[Harris and Pettitt(2014)Harris, and Pettitt]{Pettitt:PNAS}
Harris,~R.~C.; Pettitt,~B.~M. Effects of geometry and chemistry on hydrophobic
  solvation. \emph{Proc. Natl. Acad. Sci. U.S.A.} \textbf{2014}, \emph{111},
  14681--14686\relax
\mciteBstWouldAddEndPuncttrue
\mciteSetBstMidEndSepPunct{\mcitedefaultmidpunct}
{\mcitedefaultendpunct}{\mcitedefaultseppunct}\relax
\EndOfBibitem
\bibitem[Kokubo \latin{et~al.}(2013)Kokubo, Harris, Asthagiri, and
  Pettitt]{Kokubo:JPCB:2013}
Kokubo,~H.; Harris,~R.~C.; Asthagiri,~D.; Pettitt,~B.~M. Solvation Free
  Energies of Alanine Peptides: The Effect of Flexibility. \emph{J. Phys. Chem.
  B} \textbf{2013}, \emph{117}, 16428 -- 16435\relax
\mciteBstWouldAddEndPuncttrue
\mciteSetBstMidEndSepPunct{\mcitedefaultmidpunct}
{\mcitedefaultendpunct}{\mcitedefaultseppunct}\relax
\EndOfBibitem
\bibitem[Asthagiri \latin{et~al.}(2017)Asthagiri, Karandur, Tomar, and
  Pettitt]{Asthagiri_2017}
Asthagiri,~D.; Karandur,~D.; Tomar,~D.~S.; Pettitt,~B.~M. Intramolecular
  Interactions Overcome Hydration to Drive the Collapse Transition of
  Gly$_{15}$. \emph{J. Phys. Chem. B} \textbf{2017}, \emph{121}, 8078 --
  8084\relax
\mciteBstWouldAddEndPuncttrue
\mciteSetBstMidEndSepPunct{\mcitedefaultmidpunct}
{\mcitedefaultendpunct}{\mcitedefaultseppunct}\relax
\EndOfBibitem
\bibitem[Xi \latin{et~al.}(2017)Xi, Venkateshwaran, Li, Rego, Patel, and
  Garde]{Xi:PNAS:2017}
Xi,~E.; Venkateshwaran,~V.; Li,~L.; Rego,~N.; Patel,~A.~J.; Garde,~S.
  Hydrophobicity of Proteins and Nanostructured Solutes Is Governed by
  Topographical and Chemical Context. \emph{Proc. Natl. Acad. Sci. U.S.A.}
  \textbf{2017}, \emph{114}, 13345--13350\relax
\mciteBstWouldAddEndPuncttrue
\mciteSetBstMidEndSepPunct{\mcitedefaultmidpunct}
{\mcitedefaultendpunct}{\mcitedefaultseppunct}\relax
\EndOfBibitem
\bibitem[Chaudhari \latin{et~al.}(2013)Chaudhari, Holleran, Ashbaugh, and
  Pratt]{Chaudhari_2013}
Chaudhari,~M.~I.; Holleran,~S.~A.; Ashbaugh,~H.~S.; Pratt,~L.~R.
  Molecular-scale hydrophobic interactions between hard-sphere reference
  solutes are attractive and endothermic. \emph{Proc. Natl. Acad. Sci. U.S.A.}
  \textbf{2013}, \emph{110}, 20557 -- 20562\relax
\mciteBstWouldAddEndPuncttrue
\mciteSetBstMidEndSepPunct{\mcitedefaultmidpunct}
{\mcitedefaultendpunct}{\mcitedefaultseppunct}\relax
\EndOfBibitem
\bibitem[Gao \latin{et~al.}(2018)Gao, Tan, Chaudhari, Asthagiri, Pratt, Rempe,
  and Weeks]{Gao:2018}
Gao,~A.; Tan,~L.; Chaudhari,~M.~I.; Asthagiri,~D.; Pratt,~L.~R.; Rempe,~S.~B.;
  Weeks,~J.~D. Role of Solute Attractive Forces in the Atomic-Scale Theory of
  Hydrophobic Effects. \emph{J. Phys. Chem. B} \textbf{2018}, \emph{122}, 6272
  -- 6276\relax
\mciteBstWouldAddEndPuncttrue
\mciteSetBstMidEndSepPunct{\mcitedefaultmidpunct}
{\mcitedefaultendpunct}{\mcitedefaultseppunct}\relax
\EndOfBibitem
\bibitem[Sheng \latin{et~al.}(1994)Sheng, Panagiotopoulos, Kumar, and
  Szleifer]{sheng1994monte}
Sheng,~Y.~J.; Panagiotopoulos,~A.~Z.; Kumar,~S.~K.; Szleifer,~I. Monte Carlo
  calculation of phase equilibria for a bead-spring polymeric model.
  \emph{Macromolecules} \textbf{1994}, \emph{27}, 400--406\relax
\mciteBstWouldAddEndPuncttrue
\mciteSetBstMidEndSepPunct{\mcitedefaultmidpunct}
{\mcitedefaultendpunct}{\mcitedefaultseppunct}\relax
\EndOfBibitem
\bibitem[Lin \latin{et~al.}(2014)Lin, Martin, and Jayaraman]{Jayaraman}
Lin,~B.; Martin,~T.~B.; Jayaraman,~A. Decreasing Polymer Flexibility Improves
  Wetting and Dispersion of Polymer-Grafted Particles in a Chemically Identical
  Polymer Matrix. \emph{ACS Macro Lett.} \textbf{2014}, \emph{3},
  628--632\relax
\mciteBstWouldAddEndPuncttrue
\mciteSetBstMidEndSepPunct{\mcitedefaultmidpunct}
{\mcitedefaultendpunct}{\mcitedefaultseppunct}\relax
\EndOfBibitem
\bibitem[Athawale \latin{et~al.}(2007)Athawale, Goel, Ghosh, Truskett, and
  Garde]{garde07}
Athawale,~M.~V.; Goel,~G.; Ghosh,~T.; Truskett,~T.~M.; Garde,~S. Effects of
  lengthscales and attractions on the collapse of hydrophobic polymers in
  water. \emph{Proc. Nat. Acad. Sci.} \textbf{2007}, \emph{104}, 733--738\relax
\mciteBstWouldAddEndPuncttrue
\mciteSetBstMidEndSepPunct{\mcitedefaultmidpunct}
{\mcitedefaultendpunct}{\mcitedefaultseppunct}\relax
\EndOfBibitem
\bibitem[Goel \latin{et~al.}(2008)Goel, Athawale, Garde, and Truskett]{garde08}
Goel,~G.; Athawale,~M.~V.; Garde,~S.; Truskett,~T.~M. Attractions, Water
  Structure, and Thermodynamics of Hydrophobic Polymer Collapse. \emph{J. Phys.
  Chem. B} \textbf{2008}, \emph{112}, 13193--13196\relax
\mciteBstWouldAddEndPuncttrue
\mciteSetBstMidEndSepPunct{\mcitedefaultmidpunct}
{\mcitedefaultendpunct}{\mcitedefaultseppunct}\relax
\EndOfBibitem
\bibitem[Rodr{\'\i}guez-Ropero \latin{et~al.}(2015)Rodr{\'\i}guez-Ropero,
  Hajari, and van~der Vegt]{rodriguez2015mechanism}
Rodr{\'\i}guez-Ropero,~F.; Hajari,~T.; van~der Vegt,~N.~F. Mechanism of Polymer
  Collapse in Miscible Good Solvents. \emph{J. Phys. Chem. B} \textbf{2015},
  \emph{119}, 15780--15788\relax
\mciteBstWouldAddEndPuncttrue
\mciteSetBstMidEndSepPunct{\mcitedefaultmidpunct}
{\mcitedefaultendpunct}{\mcitedefaultseppunct}\relax
\EndOfBibitem
\bibitem[Cheng \latin{et~al.}(1999)Cheng, Sheu, and
  Rossky]{Rossky:1999:BioPhys.J}
Cheng,~Y.-K.; Sheu,~W.-S.; Rossky,~P.~J. Hydrophobic Hydration of Amphipathic
  Peptides. \emph{Biophys. J.} \textbf{1999}, \emph{76}, 1734 -- 1743\relax
\mciteBstWouldAddEndPuncttrue
\mciteSetBstMidEndSepPunct{\mcitedefaultmidpunct}
{\mcitedefaultendpunct}{\mcitedefaultseppunct}\relax
\EndOfBibitem
\bibitem[Shell(2010)]{Shell:2010:MS}
Shell,~M.~S. A replica-exchange approach to computing peptide conformational
  free energies. \emph{Molecular Simulation} \textbf{2010}, \emph{36},
  505--515\relax
\mciteBstWouldAddEndPuncttrue
\mciteSetBstMidEndSepPunct{\mcitedefaultmidpunct}
{\mcitedefaultendpunct}{\mcitedefaultseppunct}\relax
\EndOfBibitem
\bibitem[Zerze \latin{et~al.}(2015)Zerze, Mullen, Levine, Shea, and
  Mittal]{Shea:2015:Langmuir}
Zerze,~G.~H.; Mullen,~R.~G.; Levine,~Z.~A.; Shea,~J.-E.; Mittal,~J. To What
  Extent Does Surface Hydrophobicity Dictate Peptide Folding and Stability near
  Surfaces? \emph{Langmuir} \textbf{2015}, \emph{31}, 12223--12230\relax
\mciteBstWouldAddEndPuncttrue
\mciteSetBstMidEndSepPunct{\mcitedefaultmidpunct}
{\mcitedefaultendpunct}{\mcitedefaultseppunct}\relax
\EndOfBibitem
\bibitem[Deng \latin{et~al.}(2015)Deng, Zhang, and Levy]{Ala2-Levy-2015}
Deng,~N.; Zhang,~B.~W.; Levy,~R.~M. Connecting {{Free Energy Surfaces}} in
  {{Implicit}} and {{Explicit Solvent}}: An {{Efficient Method}} to {{Compute
  Conformational}} and {{Solvation Free Energies}}. \emph{J. Chem. Theory
  Comp.} \textbf{2015}, \emph{11}, 2868--2878\relax
\mciteBstWouldAddEndPuncttrue
\mciteSetBstMidEndSepPunct{\mcitedefaultmidpunct}
{\mcitedefaultendpunct}{\mcitedefaultseppunct}\relax
\EndOfBibitem
\bibitem[Chekmarev \latin{et~al.}(2004)Chekmarev, Ishida, and
  Levy]{Ala2-Levy-2004}
Chekmarev,~D.~S.; Ishida,~T.; Levy,~R.~M. Long-{{Time Conformational
  Transitions}} of {{Alanine Dipeptide}} in {{Aqueous Solution}}:\,
  {{Continuous}} and {{Discrete}}-{{State Kinetic Models}}. \emph{J. Phys.
  Chem. B} \textbf{2004}, \emph{108}, 19487--19495\relax
\mciteBstWouldAddEndPuncttrue
\mciteSetBstMidEndSepPunct{\mcitedefaultmidpunct}
{\mcitedefaultendpunct}{\mcitedefaultseppunct}\relax
\EndOfBibitem
\bibitem[Anand \latin{et~al.}(2010)Anand, Sharma, Dutta, Kumar, and
  Belfort]{anand2010conformational}
Anand,~G.; Sharma,~S.; Dutta,~A.~K.; Kumar,~S.~K.; Belfort,~G. Conformational
  transitions of adsorbed proteins on surfaces of varying polarity.
  \emph{Langmuir} \textbf{2010}, \emph{26}, 10803--10811\relax
\mciteBstWouldAddEndPuncttrue
\mciteSetBstMidEndSepPunct{\mcitedefaultmidpunct}
{\mcitedefaultendpunct}{\mcitedefaultseppunct}\relax
\EndOfBibitem
\bibitem[Verlet and Weis(1972)Verlet, and Weis]{VerletWeis}
Verlet,~L.; Weis,~J.-J. Equilibrium Theory of Simple Liquids. \emph{Phys. Rev.
  A} \textbf{1972}, \emph{5}, 939--952\relax
\mciteBstWouldAddEndPuncttrue
\mciteSetBstMidEndSepPunct{\mcitedefaultmidpunct}
{\mcitedefaultendpunct}{\mcitedefaultseppunct}\relax
\EndOfBibitem
\bibitem[Speedy(1987)]{speedy}
Speedy,~R.~J. Diffusion in the hard sphere fluid. \emph{Mol. Phys.}
  \textbf{1987}, \emph{62}, 509--515\relax
\mciteBstWouldAddEndPuncttrue
\mciteSetBstMidEndSepPunct{\mcitedefaultmidpunct}
{\mcitedefaultendpunct}{\mcitedefaultseppunct}\relax
\EndOfBibitem
\bibitem[Andersen \latin{et~al.}(1971)Andersen, Weeks, and Chandler]{blip}
Andersen,~H.~C.; Weeks,~J.~D.; Chandler,~D. Relationship between the
  hard-sphere fluid and fluids with realistic repulsive forces. \emph{Phys.
  Rev. A} \textbf{1971}, \emph{4}, 1597--1607\relax
\mciteBstWouldAddEndPuncttrue
\mciteSetBstMidEndSepPunct{\mcitedefaultmidpunct}
{\mcitedefaultendpunct}{\mcitedefaultseppunct}\relax
\EndOfBibitem
\bibitem[Chandler(1974)]{Chandler:AccChemRes:1974}
Chandler,~D. Equilibrium Structure and Molecular Motion in Liquids. \emph{Acc.
  Chem Res.} \textbf{1974}, \emph{7}, 246--251\relax
\mciteBstWouldAddEndPuncttrue
\mciteSetBstMidEndSepPunct{\mcitedefaultmidpunct}
{\mcitedefaultendpunct}{\mcitedefaultseppunct}\relax
\EndOfBibitem
\bibitem[Sharma and Debenedetti(2012)Sharma, and
  Debenedetti]{sharma2012evaporation}
Sharma,~S.; Debenedetti,~P.~G. Evaporation rate of water in hydrophobic
  confinement. \emph{Proc. Nat. Acad. Sci.} \textbf{2012}, \emph{109},
  4365--4370\relax
\mciteBstWouldAddEndPuncttrue
\mciteSetBstMidEndSepPunct{\mcitedefaultmidpunct}
{\mcitedefaultendpunct}{\mcitedefaultseppunct}\relax
\EndOfBibitem
\bibitem[Sharma and Debenedetti(2012)Sharma, and Debenedetti]{sharma2012free}
Sharma,~S.; Debenedetti,~P.~G. Free energy barriers to evaporation of water in
  hydrophobic confinement. \emph{J. Phys. Chem. B} \textbf{2012}, \emph{116},
  13282--13289\relax
\mciteBstWouldAddEndPuncttrue
\mciteSetBstMidEndSepPunct{\mcitedefaultmidpunct}
{\mcitedefaultendpunct}{\mcitedefaultseppunct}\relax
\EndOfBibitem
\bibitem[Miller \latin{et~al.}(2007)Miller, Vanden-Eijnden, and
  Chandler]{Miller:PNAS:2007}
Miller,~T.; Vanden-Eijnden,~E.; Chandler,~D. Solvent coarse-graining and the
  string method applied to the hydrophobic collapse of a hydrated chain.
  \emph{Proc. Natl. Acad. Sci. U.S.A.} \textbf{2007}, \emph{104},
  14559--14564\relax
\mciteBstWouldAddEndPuncttrue
\mciteSetBstMidEndSepPunct{\mcitedefaultmidpunct}
{\mcitedefaultendpunct}{\mcitedefaultseppunct}\relax
\EndOfBibitem
\bibitem[Lei \latin{et~al.}(2015)Lei, Mundy, Schenter, and
  Voulgarakis]{Mundy:JCP:2015}
Lei,~H.; Mundy,~C.~J.; Schenter,~G.~K.; Voulgarakis,~N.~K. Modeling nanoscale
  hydrodynamics by smoothed dissipative particle dynamics. \emph{J. Chem.
  Phys.} \textbf{2015}, \emph{142}, 194504\relax
\mciteBstWouldAddEndPuncttrue
\mciteSetBstMidEndSepPunct{\mcitedefaultmidpunct}
{\mcitedefaultendpunct}{\mcitedefaultseppunct}\relax
\EndOfBibitem
\bibitem[Foley \latin{et~al.}(2015)Foley, Shell, and Noid]{Noid:2015:JCP}
Foley,~T.~T.; Shell,~M.~S.; Noid,~W.~G. The impact of resolution upon entropy
  and information in coarse-grained models. \emph{J. Chem. Phys.}
  \textbf{2015}, \emph{143}\relax
\mciteBstWouldAddEndPuncttrue
\mciteSetBstMidEndSepPunct{\mcitedefaultmidpunct}
{\mcitedefaultendpunct}{\mcitedefaultseppunct}\relax
\EndOfBibitem
\bibitem[Sanyal and Shell(2016)Sanyal, and Shell]{Shell:2016:JCP}
Sanyal,~T.; Shell,~M.~S. Coarse-grained models using local-density potentials
  optimized with the relative entropy: Application to implicit solvation.
  \emph{J. Chem. Phys.} \textbf{2016}, \emph{145}\relax
\mciteBstWouldAddEndPuncttrue
\mciteSetBstMidEndSepPunct{\mcitedefaultmidpunct}
{\mcitedefaultendpunct}{\mcitedefaultseppunct}\relax
\EndOfBibitem
\bibitem[Saunders \latin{et~al.}(2013)Saunders, Voth, Voth, and
  Voth]{saunders_coarse-graining_2013}
Saunders,; Voth,~G.~A.; Voth,~G.~A.; Voth,~G.~A. Coarse-graining {Methods} for
  {Computational} {Biology}. \emph{Annu. Rev. Biophysics} \textbf{2013},
  \emph{42}, 73--93\relax
\mciteBstWouldAddEndPuncttrue
\mciteSetBstMidEndSepPunct{\mcitedefaultmidpunct}
{\mcitedefaultendpunct}{\mcitedefaultseppunct}\relax
\EndOfBibitem
\end{mcitethebibliography}
\end{document}